\documentclass[fleqn,usenatbib]{mnras}

\usepackage{newtxtext,newtxmath}

\usepackage{mathtools}
\usepackage{graphicx}
\usepackage{float}
\usepackage{amsmath}
\usepackage{nccmath}
\usepackage{bm}
\usepackage{subfigure}
\graphicspath{ {figures/} }
\usepackage{braket}
\usepackage{comment}

\usepackage[T1]{fontenc}

\DeclareRobustCommand{\VAN}[3]{#2}
\let\VANthebibliography\thebibliography
\def\thebibliography{\DeclareRobustCommand{\VAN}[3]{##3}\VANthebibliography}

\makeatletter
\renewcommand*\env@matrix[1][\arraystretch]{%
  \edef\arraystretch{#1}%
  \hskip -\arraycolsep
  \let\@ifnextchar\new@ifnextchar
  \array{*\c@MaxMatrixCols c}}
\makeatother

\usepackage{graphicx}	
\usepackage{amsmath}	

\title[Autospectra recovery from crosscorrelations]{A statistical framework for recovering intensity mapping autocorrelations from  crosscorrelations}

\author[L. McBride and A. Liu]{
Lisa McBride,$^{1}$\thanks{E-mail: elizabeth.mcbride@mail.mcgill.ca}
Adrian Liu,$^{1}$
\\
$^{1}$Department of Physics and Trottier Space Institute at McGill, McGill University, Montreal, QC H3A 2T8, Canada\\
}

\pubyear{2023}

\begin{document}
\label{firstpage}
\pagerange{\pageref{firstpage}--\pageref{lastpage}}
\maketitle

\begin{abstract}
Intensity mapping experiments will soon have surveyed large swathes of the sky, providing information about the underlying matter distribution of the early universe.  The resulting maps can be used to recover statistical information, such as the power spectrum, about the measured spectral lines (for example, HI, [CII], and [OIII]).  However precise power spectrum measurements, such as the 21\,cm autocorrelation, continue to be challenged by the presence of bright foregrounds and non-trivial systematics.  By crosscorrelating different data sets, it may be possible to mitigate the effects of both foreground uncertainty and uncorrelated instrumental systematics. Beyond their own merit, crosscorrelations could also be used to recover autocorrelation information. Such a technique was proposed in \citet{Beane_2019} for recovering the 21\,cm power spectrum. Generalizing their result, we develop a statistical framework for combining multiple crosscorrelation signals in order to infer information about the corresponding autocorrelations.  We do this first within the Least Squares Estimator (LSE) framework, and show how one can derive their estimator, along with several alternative estimators. We also investigate the posterior distribution of recovered autocorrelation and associated model parameters.  We find that for certain noise regimes and cosmological signal modeling assumptions this procedure is effective at recovering autospectra from a set of crosscorrelations.  Finally, we showcase our framework in the context of several near-future line intensity mapping experiments.
\end{abstract}

\begin{keywords}
dark ages, reionization, first stars -- large-scale structure of Universe -- methods: statistical
\end{keywords}



\section{Introduction} \label{sec:intro}
Line intensity mapping (LIM) has emerged as a new modality for extracting information from the early universe. In contrast to more traditional spectroscopic surveys which rely on identifying individual objects, LIM measures the total integrated emission of spectral lines. By exchanging this resolution for both greater sensitivity and increased field of view, large volumes of the sky can be surveyed at high redshift, enabling detailed studies of large scale structure (LSS). Crucially, because intensity mapping measures spectral lines with known rest frequencies, the observing frequency can be converted into redshifts, allowing the tomographic construction of 3D maps of the universe. Various experiments (e.g.\,\citealt{HERA, CHORD, CHIME, TianLai, HIRAX, CCAT-prime, COMAP, EXCLAIM, CONCERTO, SPHEREx}) are currently or beginning to come online, and will soon yield vast quantities of new information about the state of our high-redshift universe. 

Many emission lines have garnered interest as targets for intensity mapping (for a review, see \citealt{IMreview}). While the 21\,cm line \citep{OG_21cm} has occupied center stage for some time, there has been increasing theoretical interest and observational investment in a host of other spectral lines. These include ionized carbon [CII] \citep{CII_line, gong2011intensity}, CO rotational lines \citep{breysse2016high, li2016connecting}, and a variety of oxygen lines such as [OI] and [OIII] \citep{fonseca2017cosmology}. While modelling uncertainty of their host environments still exists, spectral lines are believed to trace the overall matter density field on sufficiently large scales.  

Depending on the processes that drive emission, the intensity of a given line can be correlated or anti-correlated with the matter density field. Some lines are also highly correlated with processes involved in galaxy assembly and star formation (for example the [CII] 158 micron line has been observed to be bright in star-forming galaxies at low redshift \citealt{croxall2017origins}), and so in addition to potential insight into the matter density field, intensity maps can provide a whole host of new information about high-redshift astrophysics, including galaxy formation, early stellar populations, and the intergalactic medium (IGM; see \citet{bernal2022line} for a concise summary).

Unfortunately all spectral lines are contaminated to varying degrees by foreground emission which can dominate over the signal of interest itself. This is especially true for high-redshift 21\,cm measurements, where the foregrounds are projected to be of order \char`\~$10^5$ larger than the desired signal \citep{di2002radio}.   Therefore, bright foregrounds make a rigorous detection of the 21\,cm autocorrelation extremely challenging, especially in combination with instrumental systematics. This is a known problem, and thus a lot of work has gone into better characterization of systematics and improved methods of foreground mitigation (see \citealt{liu2020data} for an overview). Despite these difficulties, multiple intrepid efforts are currently underway \citep{theMWA, PAPER, GMRT, LOFAR, PAPER, SKA, MeerKAT, GBT, HERA}. 

 However, it may be possible to recover much of the desired information by crosscorrelating multiple LIM measurements.  One of the attractive characteristics of crosscorrelation measurements is that if the associated foregrounds and instrumental systematics are uncorrelated between lines, they will on average crosscorrelate to zero, resulting in an unbiased measurement.  Indeed, to date, the cosmological 21\,cm signal has been detected primarily via crosscorrelations; first by the Green Bank Telescope (GBT) with a selection of galaxies observed by the Sloan Digital Sky Survey (SDSS) \citep{borthakur2010gbt}, and also with the WiggleZ galaxy survey \citep{masui2013measurement}. The WiggleZ galaxy survey was again used by \citet{cunnington2022hi}, in combination with MeerKAT radio intensity maps. The CHIME experiment achieved an interferometeric milestone with their detection of the 21\,cm signal via the use of stacking of eBOSS catalogs \citep{amiri2023detection}.  Finally, a recent direct detection of the 21\,cm autocorrelation at redshifts $z=0.32$ and $z=0.44$ by MeerKAT \citep{Sourabh} presents the beginning of a new era for 21\,cm science.
 
 For the [CII] line, \citet{pullen2018search} placed constraints on emission at a redshift range of $z=2$ to $z=3.2$ by crosscorrelating Planck high frequency maps with the SDSS quasar catalog and found an excess consistent with [CII] emission. An additional constraint was placed at redshifts $z=0.35$ and $z=0.57$ with the use of spherical harmonic tomography (SHT) between FIRAS maps and the BOSS galaxy survey \citep{anderson2022constraining}.  

These additional crosscorrelations measurements are useful because foreground contamination is not unique to 21\,cm cosmology and is an issue for all LIM experiments. Therefore crosscorrelations measurements can be useful for circumventing autocorrelation issues across all spectral lines.
 Proposed line combinations include 21\,cm with [CII], CO, Ly-$\alpha$, and hydrogen deuteride \citep{Chung_2020, silva2013intensity, chang2015synergy,breysse2021mapping}, [CII] with [OIII] \citep{Padmanabhan_2020}, as well as crosscorrelating intensity maps with more traditional spectroscopic surveys \citep{villaescusa2015cross,Breysse_2019,Chung_2019, la2022prospects}. Even though continuum foregrounds such as the ones that plague hydrogen mapping experiments are less bright at the (marginally) higher frequency bands of other target spectral lines  \citep{aghanim2016planck}, this can still be an attractive pathway towards first detections. Although it must be noted that LIM experiments can suffer from interloper line contamination \citep{cheng2020phase,sun2018foreground} that does not affect 21\,cm experiments.

While crosscorrelations are attractive in their own right, it may also be possible to infer autocorrelation information from crosscorrelation measurements.  Reconstructing the 21\,cm field in this fashion would provide a complementary analysis alongside direct 21\,cm measurements, and serve as a helpful check of foreground removal techniques.  This may be particularly necessary for 21\,cm observations due to its uniquely challenging nature. Such a method has been proposed recently by \citet{Beane_2019} (B19 from hereon) as a way to sidestep these challenges.  They propose recovering the 21\,cm brightness temperature power spectrum using three crosscorrelations of neutral hydrogen, [CII], and [OIII] lines. In their work, they were able to recover the autocorrelation of their simulated 21\,cm signal to within $\sim$5\% accuracy for the majority of the region in redshift where reionization is expected to occur (between $z=7$ and $z=9$ for the scenario assumed in B19), known as the Epoch of Reionization (EOR).

Given the enticing results in B19, we ground their results in a broader statistical framework via a linearized formalism, and show that the B19 estimator can be derived in this manner. This framework allows us to generalize their technique to a broader suite of similar estimators. In addition to this more frequentist-based analysis, we also conduct a numerical Bayesian investigation of the full posterior distribution, in order to better understand the error uncertainties on such a measurement. We also examine the model dependence of our estimators by fitting simulated data sets derived from three different models in order to understand the regimes in which the linear biasing model---assumed by all of our estimators---breaks down, and how it affects our results.

The rest of this paper is structured as follows; we begin with a brief overview of the linear biasing model assumed throughout this work in Section\,\ref{sec:models}. In Section\,\ref{sec:estimators} we outline the statistical and mathematical formalism for the various estimators investigated, and define our Least Squares Estimation-based (LSE) estimator. We then set out to explore both the B19 estimator and our LSE estimator in Section\,\ref{sec:results}, using a variety of simulated data described in Section\,\ref{sec:simulations}.  A summary of the content within can be found in Section\,\ref{sec: conclusions}.

\section{Models} \label{sec:models}

\subsection{Linear Biasing Models} \label{sec: biasing}
The underlying matter density cannot be measured directly; however, on sufficiently large scales baryonic matter follows the underlying matter density. Thus, regions of strong line emission are associated with regions of higher matter density.  In this way, spectral lines are biased tracers of the underlying matter density. This coupling between the dark matter and baryonic fields allows us to exploit the wide variety of emission from the early universe.  

All the estimators discussed in this work are constructed assuming the linear biasing model for all spectral lines. In this model, the intensity $I$ of a spectral line $i$ (expressed as $I_i$) can be written in terms of the Fourier space matter overdensity $\delta(\mathbf{k})$, written here as a function of the Fourier dual $\mathbf{k}$ of the spatial position $\mathbf{x}$.
\begin{align}
\label{eq:Ibdefinition}
I_i(\mathbf{k}) = \braket{I_i} b_i \delta(\mathbf{k})
\end{align}
where $b_i$ is the bias factor, and $ \braket{I_i}$ is the mean intensity of the line.

Given an intensity field, one can calculate the resulting power spectrum, or autocorrelation,
\begin{align}
P_{ii}(k,z) =\left[ \braket{I_i}(z) b_i(z) \right]^2 P_m (k,z)
\end{align}
which is similarly related to the matter power spectrum, $P_m(k,z)$.

Alternatively, one could combine two differing spectral lines, for instance $i$ and $j$, and calculate their cross-power spectrum, or crosscorrelation,
\begin{align} \label{eq:linearbias}
P_{ij} = \braket{I_i} \braket{I_j} b_i b_j P_m
\end{align}
where we have suppressed the wavenumber and redshift dependence for brevity.

\subsection{Physical Models} \label{sec: physical}
Below, we provide a brief physical description of each of our target spectral lines, along with our assumed values of each line's predicted intensities.  Because we are more interested in the statistical properties of such a measurement and less interested in the astrophysical modeling of line intensities, we use a shorthand expression for the constant factors in Equation~\eqref{eq:Ibdefinition}. For our analysis, we defined the intensity-weighted bias factor,
\begin{align}
 \beta_i \equiv b_i \braket{I_i}
 \end{align}
which we then set to be consistent with current astrophysical models. While the luminosity functions that govern the shape of the intensity are still not well known at high redshift, we note that the qualitative results of our analysis are not affected by an overall scaling of the signals (as long as the noise is similarly scaled). Additionally, our analysis is done on a per-$k$ basis, and therefore will also not be affected by potential scale-dependent biases. 

Below we briefly describe the three spectral lines used in our analysis. For more motivation of the use of these target lines, plus additional emission sources of interest for LIM, see \citet{carilli2013cool}.

\subsubsection{21\,cm}
Due to hyperfine splitting, neutral hydrogen can undergo a spin flip transition that releases or absorbs a radio frequency photon with a rest wavelength of 21~cm. The rarity of this transition is balanced by the sheer volume of neutral hydrogen in the universe.  Since other sources of spectral emission require star formation, it is the only line observable from the highest redshifts, before galaxy assembly, in an epoch referred to as the Dark Ages.  

However, a large quantity of neutral hydrogen also exists at the redshifts associated with Cosmic Dawn, when other spectral lines are predicted to appear due to star formation and the subsequent creation of dust and molecular gas.  In addition, the universe undergoes a critical phase transition during the EOR where the intergalactic medium (IGM) becomes fully ionized. This makes it an attractive target line across a broad range of redshifts as it probes all epochs of interest in cosmology, from the pristine primordial power spectrum, through structure formation, to the  state of our universe as it is today \citep{scott199021,furlanetto2009cosmology,zaldarriaga200421}. For a review of the current landscape of 21\,cm measurements, see \citet{liu2020data}.

Consistent with conventions in radio astronomy, we will use the brightness temperature, $T_b$, which is often used as a proxy for the intensity.  In the Rayleigh-Jeans limit ($h \nu \ll kT$), the former can be expressed in terms of the latter as
\begin{align}
   T_b = \frac{I_{\nu} c^2}{2k_b \nu^2}.
\end{align}
where $I_{\nu}$ is spectral intensity at a frequency $\nu$, $c$ is the speed of light, and $k_B$ is the Boltzmann constant. For our analysis the average brightness temperature is calculated from the \texttt{zreion} simulation discussed in Section\,\ref{sec:bt} to be $\braket{T_{\text{21cm}}} = {\beta}_i^{2} \ = 2.06 \times 10^3$\,Jy/sr. This value is then used to fix the brightness temperature for the additional data sets.
  
\subsubsection{[CII]}
The atomic fine structure line of singly ionized carbon [CII], with rest wavelength  $\lambda_{\mathrm{rest}}=158\,\mu$m, plays an important role in the cooling of gas in the interstellar medium and is thus strongly associated with star-forming regions.  At low redshift it is known to be one of the brightest emission lines \citep{croxall2017origins, lapham2017herschel, smith2016spatially}.  Because of its connection to star formation, [CII] is an attractive target candidate as it should be highly correlated with star forming galaxies.  However, there exists some uncertainty of the strength of the [CII] line at cosmological redshifts.  Recent observations have suggested that it is weaker than would be predicted from extrapolating based on local star-forming galaxies \citep{laporte2019absence}.  Predictions for both the luminosity function, and amplitude of the power spectrum are model dependent \citep{karoumpis2022cii, lagache2017cii} as is the range of proposed strength of the signal, with indications \citep{carniani2020missing} that the [CII] luminosity at redshifts $z > 6$ may be lower than the local value \citep{Dumitru_2019}.

 For the [CII] line, we set  $\braket{I_{\text{CII}}} = 1.1 \times 10^3$\,Jy/sr. This is derived from the models contained within \citet{CII_intensity}, and is consistent with the estimates used in B19, although it is possible that this model is overly optimistic \citep{Chung_2020}.

\subsubsection{[OIII]}
Oxygen has several features of interest in the far-infrared (FIR). Two potential targets are the fine structure lines [OIII] 52 $\mu$m and 88 $\mu$m. These transitions play an important role in cooling molecular gas, and therefore by extension star formation \citep{schimek2023high, suzuki2016iii}.

Even less is known about the [OIII] line at high redshifts compared to the previous two lines. However, galaxies have been observed to have bright [OIII] emission at redshift $z= 2.8076$ by \citet{ferkinhoff2010first}. Additionally, \citep{ALMA_OIII} has measured luminosity ratios of [OIII]/[CII] of at least greater than unity, and thus a conservative estimate of the [OIII] line would be to match the same order of magnitude for intensity, as a lower bound. We adopt this approach, setting the average specific intensity to be $\braket{I_{\text{OIII}}} = 1.0 \times 10^3$\,Jy/sr.

\section{Estimators} \label{sec:estimators}
Taking the reasonableness of the linear biasing model as an axiom, we now turn our attention to the construction of a series of estimators built to recover the autocorrelation of a single line, given a set of crosscorrelation-only measurements.  We would like to recover the associated bias factors and the matter power spectrum for the lines measured.  In addition, we would like to infer the value of a single autocorrelation (for example the 21~cm power spectrum) from the estimated model parameters via a reconstruction based on the linear biasing model.

A few different estimators are described in detail below. In addition to these analytic estimators, is it possible to derive best-fit values from the full posterior function, discussed in Section\,\ref{sec:Bayes!}. 

\subsection{Beane et al. Estimator} \label{sec: Beane}
The B19 estimator does not include any measurements of the autocorrelation and instead treats the autocorrelation signal as the parameter to be estimated.  Since the intensity of each spectral line is a simple scalar product of the bias factors and the matter power spectrum in the linear biasing model, one can reconstruct a single autocorrelation via a particular arrangement of the crosscorrelations such that the undesired bias factors drop out,
\begin{align} \label{eq:B19}
\hat{P}_{ii, \text{B19}} &= R_{ijk} \frac{P_{ij} P_{ik}}{P_{jk}}
\end{align} 
where the hat in Equation\,\ref{eq:B19} indicates an estimator. Here the relative correlation between all three lines is bundled into the trinary crosscorrelation coefficient, $R_{ijk}$, and is assumed to be of order unity on the large scales assumed in this work, as well as in B19.  However, we challenge this assumption in Section\,\ref{sec:results}.

\subsection{Least Squares Estimators} \label{sec:LSEsubsection}
In choosing an estimator, a basic wish list would be to have a) the smallest possible errors on the recovered parameters, and b) it be unbiased.  These two requirements lead us quite naturally to consider the Least Squares Estimator (LSE).

The LSE method is a workhorse of frequentist techniques for recovering the best-fit values of a set of model parameters given a set of observations.  Under certain conditions the LSE is optimal in that it yields an unbiased fit, with the smallest possible errors.  Its properties are well-understood, which makes it a compelling starting point for building a suite of estimators.

A data vector, $\mathbf{d}$, described by a linear model, $\mathbf{m}\equiv \mathbf{Ax}$, can be written in matrix form as
\begin{align}
    \mathbf{d} &= \mathbf{Ax} + \mathbf{n}
\end{align}
where $\mathbf{x}$ is a vector of the parameters which are related to the data via the matrix $\mathbf{A}$, and $\mathbf{n}$ is the noise vector of a particular realization of the observational noise. Inverting this expression gives an estimate of the values of the parameters, $\hat{\mathbf{x}}$,
\begin{align} \label{eq:LSE}
 \hat{\mathbf{x}} &= [\mathbf{A}^T \mathbf{N}^{-1} \mathbf{A}]^{-1} \mathbf{A}^T \mathbf{N}^{-1} \mathbf{d}
\end{align}
where the noise covariance matrix, $\mathbf{N}$, is formed from calculating the ensemble average of the outer product of the noise vector (i.e., $\mathbf{N}\equiv \braket{\mathbf{nn}^T}$ in the case of mean-zero noise) and fully characterizes the observational noise in the case that noise is Gaussian.
We apply this formalism to a few cases below.

\subsubsection{Logarithmic Estimator for three crosscorrelations and a bias prior} \label{sec:3cc1p}
Using B19 as a reference, the minimal set of observations necessary to recover a sole autocorrelation, $P_{ii}$, are three crosscorrelations: $P_{ij}$, $P_{jk}$, $P_{ik}$.  Other choices for additional observations exist, and we will discuss a few options and their validity in following sections. 

If we are also interested in recovering the three bias factors and matter power spectrum, we must add an additional data point lest the system of equations be undetermined.  This is due to the degeneracy between the biases and the matter power spectrum, where the data remain the same if one multiplies all the biases by a constant factor while simultaneously dividing the matter power spectrum by the square of this factor.\footnote{The relevant system of equations in this section are in fact mathematically analogous to those that describe the calibration of a radio interferometer through redundant information \citep{Liu_2010} This degeneracy is mathematically identical to the overall gain degeneracy in redundant-baseline calibration \citep{dillon2020redundant, liu2020data}.}

As a way to break this degeneracy, we choose to include an additional measurement on a single bias factor, $\beta_i$. For the assumptions here this is mathematically equivalent to including a Gaussian prior on $\beta_i$.  We will also see that with our current choice of data, the recovered autocorrelation is largely insensitive to the choice of prior.

Our set of models are then
\begin{align} \label{eq:data}
P_{ij} &= \beta_i \beta_j P_m + n_{ij} \nonumber \\
P_{jk} &= \beta_j \beta_k P_m + n_{jk} \nonumber \\
P_{ki} &= \beta_k \beta_i P_m + n_{ki} \nonumber \\
\beta_0 &= \beta_i + n_{b_0}
\end{align}
where we have again suppressed the $k$-dependence for clarity.
Following the general procedure outlined in \cite{Liu_2010}, we linearize the signals; first by rewriting the bias factors as exponentials.  Then, the equations become
\begin{align}
P_{ij} &= e^{\eta_i} e^{\eta_j} P_m (k) + n_{ij} \nonumber \\
P_{jk} &= e^{\eta_j} e^{\eta_k}  P_m (k) + n_{jk} \nonumber \\
P_{ki} &= e^{\eta_k} e^{\eta_i}  P_m (k) + n_{ki} \nonumber \\
\beta_0 &= e^{\eta_i} + n_{b_0} .
\end{align}

We define the quantity
\begin{equation} \label{eq:wdef}
w_{\mu \nu} \equiv \ln\left( 1 + \frac{n_{\mu \nu}}{e^{\eta_{\mu}} e^{\eta_{\nu}} P_m (k)} \right)
\end{equation}
with $\mu,\nu \in \{i,j,k \}$ for the noise term in each crosscorrelation, and a similar expression for the uncertainty on the prior. This allows us to rewrite our equations in the more compact form,
\begin{align} \label{eq:system of eqs}
\ln P_{ij} &= \eta_i + \eta_j + \ln[P_m (k)] + w_{ij} \nonumber \\
\ln P_{jk} &= \eta_j + \eta_k + \ln[P_m (k)] + w_{jk} \nonumber \\
\ln P_{ki} &= \eta_k + \eta_i + \ln[P_m (k)] + w_{ki} \nonumber \\
\ln{\beta_0} &= \eta_i + w_{b_0}
\end{align}
We then have a linear system of equations which can be expressed using matrix notation.  We identify the LHS as the data,
\begin{align}
    \mathbf{d} \equiv \begin{pmatrix}
       \ln P_{ij} \\
       \ln P_{jk} \\
       \ln P_{ki} \\
       \ln{\beta_0}
    \end{pmatrix},
\end{align}
and the RHS as the model,
\begin{align}
    \mathbf{Ax} \equiv \begin{pmatrix}
       \eta_i + \eta_j + \ln[P_m (k)] \\
       \eta_j + \eta_k + \ln[P_m (k)] \\
       \eta_k + \eta_i + \ln[P_m (k)] \\
       \eta_i
    \end{pmatrix},
\end{align}
plus a noise realization,
\begin{align}
    \mathbf{n} = \begin{pmatrix}
        w_{ij} \\
        w_{jk} \\
        w_{ki} \\
        w_{b_0}
    \end{pmatrix}.
\end{align}
Utilising the least squares formalism [namely, Equation~\eqref{eq:LSE}] we then calculate the estimated values of the logarithm of the parameters.  We then exponentiate each element to find estimators $\hat{\beta}_i$, $\hat{\beta}_j$, $\hat{\beta}_k$, and $\hat{P}_m$ for our original parameters, giving
\begin{align} \label{eq:xhat}
\begin{pmatrix}
\hat{\beta}_i \\
\hat{\beta}_j \\
\hat{\beta}_k \\
\hat{P}_m \\
\end{pmatrix} = \begin{pmatrix}[1.25]
\beta_0 \\
\beta_0 P_{jk} / P_{ki} \\
\beta_0 P_{jk} / P_{ij} \\
\frac{P_{ij} P_{ki}}{\beta_0^2 P_{jk}}  \\
\end{pmatrix}.
\end{align}
From here we combine the estimators for $\beta_i$ and $P_m$ to construct an estimator for $P_{21}$. If we identify the $i$-th line with $21$\,cm, and $j,k$ with [CII] and [OIII], respectively, then
\begin{align} \label{eq:P_21}
\hat{P}_{21} &= \hat{\beta}_i^2 \hat{P}_m = \beta_0^2 \frac{P_{ij} P_{ki}}{\beta_0^2 P_{jk}} = \frac{P_{21\textrm{cm/CII}} P_{21\textrm{cm/OIII}}}{P_{\textrm{CII/OIII}}} 
\end{align}
where $P_{\textrm{CII/OIII}}$ denotes the crosscorrelation of [CII] and [OIII], and so on, meaning we have assigned specific spectral lines to each index. The above expression is then equivalent to Equation\,\ref{eq:B19}.  In this way, the B19 estimator emerges naturally from the LSE formalism.

For the LSE estimator described by Equation\,\ref{eq:xhat} it is notable that the inferred parameters for this set of observations are independent of the form of the noise matrix, $\mathbf{N}$, which has not yet been discussed. Thus, even in the case of highly correlated noise this estimator is unchanged. Intuitively, this minimal set of data does not have room to favor one measurement over the other, as there is no other additional information from which to draw. One is driven to the B19 estimator regardless of the form of $\mathbf{N}$. For instance, it is not necessary to assume that $\mathbf{N}$ is diagonal. This is important because the appropriateness of a diagonal $\mathbf{N}$ depends on the signal-to-noise ratio of one's measurements. Casting aside for now the difference in noise statistics between the original measurements of the cross power spectra in Equation~\eqref{eq:data} and the logarithmic version in Equation~\eqref{eq:system of eqs} (a caveat that we address later in this section), consider the noise covariance between our cross power spectra in the regime where instrumental noise dominates over cosmic variance. As a first guess, one might suspect that there might be a correlation between the noise in $P_{ij}$ and $P_{jk}$ since they both have the $j$th line in common. This is in fact not the case. To see this, imagine that $n_{ij} \sim \delta T_{i} \delta T_{j}$, where $\delta T_{i}$ is the map-level (as opposed to power spectrum) noise realization in the $i$th line. The noise covariance between $P_{ij}$ and $P_{jk}$ then goes like
\begin{equation}
    \langle n_{ij} n_{jk} \rangle \sim \langle \delta T_{i} \delta T_{j} \delta T_{j} \delta T_{k} \rangle = \langle \delta T_{i} \rangle \langle \delta T_{j}^2 \rangle  \langle \delta T_{k} \rangle = 0,
\end{equation}
where we have assumed that the measurement of each individual line is independent.\footnote{Readers that find it helpful to think in the analogy with radio interferometry may recognize this argument. It is mathematically identical to the argument for why different baselines of an interferometer have independent noise realizations even if they share a common antenna, so long as enough time samples are accumulated in the correlator per integration.} The situation changes in the high signal-to-noise limit when cosmic variance dominates. In such a regime, the variance in one's measurement is dominated by the variance in the underlying matter density field itself, which affects all the lines in the same way. The key message here is that with a minimal set of measurements, $\mathbf{N}$ does not affect how we weight our relative data.

In addition to (potentially) informing the way that data is weighted (or not, as is our case), the $\mathbf{N}$ matrix also serves as crucial input within the LSE in determining the final error covariances $\boldsymbol{\Sigma}$ on parameters. In principle, this is simply given by $\boldsymbol{\Sigma} = [\mathbf{A}^T \mathbf{N}^{-1} \mathbf{A}]^{-1}$. However, in linearizing our expressions we have changed the statistics of the noise in a non-trivial way.  The projected measurement errors derived from the Least Squares formalism are only approximately correct at reasonably high signal-to-noise when the logarithm in $w_{\mu \nu}$ can be Taylor expanded (as we show in Appendix\,\ref{sec:N}). For the rest of this subsection, we will work in this limit.  We also assume that the prior on the bias $\beta_0$ is Gaussian in the linearized form.

In this limit, the new observational errors are given by an approximate noise matrix,
\begin{align} \label{eq:Ntilde}
\mathbf{\tilde{N}} = \begin{pmatrix} 
\frac{\sigma_{ij}^2}{P_{ij}^2} & 0 & 0 & 0 \\
0 & \frac{\sigma_{jk}^2}{P_{jk}^2} & 0 & 0 \\
0 & 0 & \frac{\sigma_{ki}^2}{P_{ki}^2} & 0 \\
0 & 0 & 0 & \frac{\sigma_{\beta_i}^2}{\beta_{0}^2}
\end{pmatrix}.
\end{align} 
From here we can recover the estimated variance on the original parameters (see Appendix\,\ref{sec:delog}):
\begin{align}
\begin{pmatrix}
\sigma_{\beta_i}^2 \\
\sigma_{\beta_j}^2 \\
\sigma_{\beta_k}^2 \\
\sigma_{P_m}^2 \\
\end{pmatrix} =\begin{pmatrix}
\sigma_{\beta_i}^2  \\
 \left(\frac{\sigma_{\beta_i}^2}{\beta_i^2} + \frac{\sigma_{ik}^2}{P_{ik}^2} + \frac{\sigma_{jk}^2}{P_{jk}^2}\right) \beta_j^2 \\
 \left(\frac{\sigma_{\beta_i}^2}{\beta_i^2} + \frac{\sigma_{ij}^2}{P_{ij}^2} + \frac{\sigma_{jk}^2}{P_{jk}^2}\right) \beta_k^2 \\
\left(4\frac{\sigma_{\beta_i}^2}{\beta_i^2} + \frac{\sigma_{ij}^2}{P_{ij}^2} + \frac{\sigma_{ik}^2}{P_{ik}^2} + \frac{\sigma_{jk}^2}{P_{jk}^2}\right)P_m^2 \\
   \end{pmatrix},
\end{align}
where $\sigma_{\beta_i}^2$, $\sigma_{\beta_j}^2$, $\sigma_{\beta_k}^2$, and $\sigma_{P_m}^2$ are the variances on $\beta_i$, $\beta_j$, $\beta_k$, and $P_m$, respectively. Using the procedure described in Appendix~\ref{sec:delog}, we find the propagated variance $\sigma^2_{P_{ii}}$ on $P_{ii}$ to be
\begin{align} \label{eq:errorpropPii}
\sigma^2_{P_{ii}}  &= \left( \frac{\sigma_{ij}^2}{P_{ij}^2} + \frac{\sigma_{jk}^2}{P_{jk}^2} + \frac{\sigma_{ik}^2}{P_{ik}^2} \right) P_{ii}^2.
\end{align}

\subsubsection{Logarithmic Estimators for three crosscorrelations, one autocorrelation, and a bias prior} \label{sec:3cc1ac1p}
Once developed, the above formalism can be applied to a multitude of situations.  Another application of interest is when there is additional information about the parameters in the form of an autocorrelation. While we expect any autocorrelation measurement to be the limiting source of precision, let us also consider a modified version of the above estimator, where the autocorrelation is included. Adding the autocorrelation measurement gives us a new set of equations,
\begin{align} \label{eq:1a3c1b}
P_{ii} &= \beta_i^2 P_m (k) \\
P_{ij} &= \beta_i \beta_j P_m (k) \nonumber \\
P_{jk} &= \beta_j \beta_k P_m (k) \nonumber \\
P_{ki} &= \beta_k \beta_i P_m (k) \nonumber \\
\beta_0 &= \beta_i .\nonumber
\end{align}
Working in the aforementioned high-signal-to-noise regime where we have approximately Gaussian noise realizations such as $\tilde{n}_{ij}$, our linearized equations are
\begin{align}
\ln{P_{ii}} &= 2 \eta_i + \ln{[P_m (k)]} + \tilde{n}_{ii} \\
\ln{P_{ij}} &= \eta_i + \eta_j + \ln{[P_m (k)]} + \tilde{n}_{ij} \nonumber \\
\ln{P_{jk}} &= \eta_j + \eta_k  + \ln{[P_m (k)]} + \tilde{n}_{jk} \nonumber \\
\ln{P_{ki}} &= \eta_k + \eta_i  + \ln{[P_m (k)]} + \tilde{n}_{ki} \nonumber \\
\ln{b_0} &= \eta_i \nonumber + \tilde{n}_{b_i}.
\end{align}
Applying the LSE formalism of Equation\,\eqref{eq:LSE} gives us the estimator for the autocorrelation line given this new set of observations:
\begin{align} \label{eq:productPii}
\hat{P}_{ii} &= P_{ii}^{1 - \varepsilon} \left( \frac{P_{ij} P_{ki} }{P_{jk}} \right)^{\varepsilon} \\
&= P_{ii}^{1 - \varepsilon} \hat{P}_{ii,\text{B19}}^{\varepsilon} \nonumber
\end{align}
where $\hat{P}_{ii,\text{B19}}$ is the B19 estimator from Section~\ref{sec: Beane} and $\varepsilon$ is defined as
\begin{align}
\varepsilon^{-1} \equiv 1+ \frac{\tilde{\sigma}_{ij}^2 + \tilde{\sigma}_{jk}^2 + \tilde{\sigma}_{ki}^2}{\tilde{\sigma}_{ii}^2}.
\end{align}
Here, we have defined the signal-weighted noise $\tilde{\sigma}_{\mu\nu} \equiv \sigma_{\mu\nu} / P_{\mu\nu}$ for conceptual clarity. From the suggestive notation it is clear that this estimator is weighting two different modalities of recovering the autocorrelation. Since the estimator has additional information over the previous estimator (which had the minimum amount of data to solve for the number of parameters), it can be pickier about which data to prioritize.  We can investigate the consequence of this by examining the error on the autocorrelation measurement and taking some illustrative limits.

If $\tilde{\sigma}_{\text{ii}}=0$, then we have measured the autocorrelation perfectly, in which case $\varepsilon \to 0$ and our estimator becomes $\hat{P}_{\text{ii}} = P_{\text{ii}}$. This is just the statement that the additional measurements are superfluous and the estimated autocorrelation is equal to the (perfect) measurement.  If, on the other hand, we do not trust the autocorrelation measurement at all, then $\tilde{\sigma}_{\text{ii}} \to \infty $ and $\varepsilon \to 1$, meaning our estimator becomes $\hat{P}_{\text{ii}} = P_{\text{ij}} P_{\text{ki}} /{P_{\text{jk}}}$, our previous result.  This again makes intuitive sense as we are essentially throwing away the new information we had added and reverting to our earlier, diminished data set.  In the intermediate case, the estimator balances the usage of both sets, downweighting each contribution by the uncertainties of the data. In fact, the logarithm of Equation~\eqref{eq:productPii} can be written as
\begin{eqnarray}
    \ln{\hat{P}_{ii}} &=& \left[ \frac{\tilde{\sigma}_{ii}^{-2}}{\left(\tilde{\sigma}_{ij}^2 + \tilde{\sigma}_{jk}^2 + \tilde{\sigma}_{ki}^2\right)^{-1} 
    + \tilde{\sigma}_{ii}^{-2}} \right] \ln{P_{ii}} \nonumber \\
    && +\left[ \frac{\left(\tilde{\sigma}_{ij}^2 + \tilde{\sigma}_{jk}^2 + \tilde{\sigma}_{ki}^2\right)^{-1} }{\left(\tilde{\sigma}_{ij}^2 + \tilde{\sigma}_{jk}^2 + \tilde{\sigma}_{ki}^2\right)^{-1} 
    + \tilde{\sigma}_{ii}^{-2}} \right]  \ln{\hat{P}_{ii,\text{B19}}},
\end{eqnarray}
demonstrating that our final estimator is essentially an inverse variance weighting of the logarithm of the individual estimators.
\subsubsection{Additional estimators}
The above formalism can incorporate as many lines as are sufficiently described by the linear bias model.  As a last example, we consider adding an additional line, indexed by $l$, to our data set which now expands to include the additional possible crosscorrelations $\mathbf{d}=\{P_{ij}, P_{jk}, P_{ik}, P_{il}, P_{jl}, P_{kl}, \beta_0\}$.  We then apply our LSE procedure and combine our estimators of $P_m$ and $\beta_i$ to form an estimator of the autocorrelation of line $i$, as above.

In order to succinctly state our result, we first define a generalized B19 estimator for arbitrary indices as
\begin{align}
\hat{P}_{\text{B19}, ijk} = \frac{P_{ij} P_{ik}}{P_{jk}}
\end{align}
and a higher-order version of the B19 estimator that takes four lines,
\begin{align}
\hat{P}_{\text{B19}, ijkl} = \frac{P_{ij}^2 P_{kl}}{P_{jk} P_{jl}}.
\end{align}
The estimator for the autocorrelation then becomes
\begin{align}
 \ln{\hat{P}_{21\mathrm{cm}}} &= \frac{\tilde{\sigma}_{jl}^2 \tilde{\sigma}_{il}^2 + \tilde{\sigma}_{kl}^2 \tilde{\sigma}_{il}^2 + \tilde{\sigma}_{kl}^2 \tilde{\sigma}_{jl}^2}{\Xi^2} 
\ln{\left(\hat{P}_{\text{B19}, ijk} \right)}  \nonumber \\
&\quad\quad 
+ \frac{\tilde{\sigma}_{kl}^2 \tilde{\sigma}_{jk}^2 + \tilde{\sigma}_{ik}^2 \tilde{\sigma}_{jk}^2 + \tilde{\sigma}_{ik}^2 \tilde{\sigma}_{kl}^2}{\Xi^2}
\ln{\left(\hat{P}_{\text{B19}, ijl} \right)} \nonumber \\
&\quad\quad  
+ \frac{\tilde{\sigma}_{ij}^2 \tilde{\sigma}_{jk}^2 + \tilde{\sigma}_{jl}^2 \tilde{\sigma}_{jk}^2 + \tilde{\sigma}_{ij}^2 \tilde{\sigma}_{jl}^2}{\Xi^2}
\ln{ \left( \hat{P}_{\text{B19}, ikl} \right)} \nonumber \\
&\quad\quad 
 + \frac{\tilde{\sigma}_{ik}^2 \tilde{\sigma}_{il}^2}{\Xi^2}\ln{\left(\hat{P}_{\text{B19}, ijkl} \right)} + \frac{\tilde{\sigma}_{ij}^2 \tilde{\sigma}_{il}^2}{\Xi^2}\ln{\left(\hat{P}_{\text{B19}, ikjl}\right)} \nonumber \\
&\quad\quad + \frac{\tilde{\sigma}_{ij}^2 \tilde{\sigma}_{ik}^2}{\Xi^2} \ln{\left(\hat{P}_{\text{B19}, iljk} \right)}, 
\end{align}
where
\begin{equation}
    \Xi^2 \equiv (\tilde{\sigma}_{ij}^2 + \tilde{\sigma}_{kl}^2) (\tilde{\sigma}_{ik}^2 + \tilde{\sigma}_{il}^2 + \tilde{\sigma}_{jk}^2 + \tilde{\sigma}_{jl}^2) + (\tilde{\sigma}_{ik}^2 + \tilde{\sigma}_{jl}^2)(\tilde{\sigma}_{il}^2 + \tilde{\sigma}_{jk}^2).
\end{equation}
In this form, we can see that the estimator is constructed from all possible combinations of the B19 estimator and its higher-order versions.  For the three possible B19 permutations, each combination appears three times, weighted by the uncertainty associated with a pair of observations. As in the similar estimator of Section~\ref{sec:3cc1p}, it is independent of the bias prior.

In the following sections, we will not explore the four-line estimator any further. The message here (as well as in previous subsections) is simply that a wide class of B19-like estimators can be constructed. Another potential generalization is to solve for the autocorrelation across all $k$ bins simultaneously (perhaps imposing constancy or a simple parametric form in the bias). We leave such generalizations to future work.

\subsection{Bayesian framework} \label{sec:Bayes!}
In addition to the frequentist estimators outlined above, we also investigate recovering parameters from a Bayesian perspective.  This is often done by examining the posterior probability distribution, or posterior.  The posterior is
\begin{align}
    p(\mathbf{x} | \mathbf{d}) = \frac{p(\mathbf{d} | \mathbf{x}) p(\mathbf{x})}{p(\mathbf{d})}
\end{align}
where the numerator is the product of the likelihood $p(\mathbf{d} | \mathbf{x})$ and the prior $p(\mathbf{x})$, and the denominator is the evidence, $p(\mathbf{d})$.  In this case we will use the same data set as the previous LSE, just prior to taking the logarithm.  Our data $\mathbf{d}$ is then the left hand side of Equation\,\ref{eq:data}. Similarly our model $\mathbf{m}(\mathbf{x})$, characterized by the parameters $\mathbf{x}$, is given by the RHS of the same set of expressions (minus the noise term), with each line corresponding to a component of $\mathbf{m}(\mathbf{x})$. 

We assume a Gaussian likelihood such that
\begin{align}
    \ln{p(\mathbf{d} | \mathbf{x})} \propto -\frac{1}{2} [\mathbf{d}-\mathbf{m}(\mathbf{x})]^{T} \mathbf{N}^{-1} [\mathbf{d}-\mathbf{m}(\mathbf{x})].
\end{align}

Since all our parameters are required to be positive for physical reasons, we apply a positivity prior to all parameters.  We additionally assume a Gaussian prior on the bias, $\beta_i$, of the autocorrelation line $P_{\mathrm{ii}}$, centered on the true value and with a standard deviation equivalent to a 10\% uncertainty.

Using an Markov Chain Monte Carlo (MCMC) algorithm, we numerically sample different sets of parameter values in order to quantify the above posterior distribution.  This gives us a series of MCMC chains, each composed of $n_{\text{steps}}$ number of samples where each sample is a set of parameter values drawn from the posterior. We are also interested in recovering the 21\,cm line autocorrelation from our set of crosscorrelations, which would provide a direct comparison to the B19 estimator. We construct the autocorrelation posterior distribution numerically by forming the autocorrelation $P_{ii}$ as defined in the first equation of Equation~\eqref{eq:1a3c1b} from each sample of the MCMC chains.

Sampling a posterior distribution allows for a subtler description of uncertainties than simply quoting a best-fit value for an estimator and its associated error bar. This is particularly important given the form of the estimators in this paper (and in B19). Since \emph{ratios} of the random variables are involved, the posterior distribution of the estimated autocorrelations can in principle be highly non-Gaussian. Our expressions for error bars (from previous subsections) should therefore be interpreted cautiously, as they implicitly assume that all measurements have high signal-to-noise. When this assumption is broken, our simple expressions can fail badly. One important scenario where this turns out to be the case is when the cross-correlation spectrum that appears in the denominator of a B19-style estimator has large errors. For example, suppose that we have high signal-to-noise data from $21\,\textrm{cm}$ measurements and relatively low signal-to-noise measurements from [CII] and [OIII], but are worried about systematics in the $21\,\textrm{cm}$ observations. Since cross-correlation errors go like the geometric mean of the map errors of each constituent line, this means that $P_{21\textrm{cm/CII}}$ and $P_{21\textrm{cm/OIII}}$ will have relatively small errors whereas $P_{\textrm{CII/OIII}}$ will possess large errors. If one uses Equation~\eqref{eq:P_21} to estimate the $21\,\textrm{cm}$ autocorrelation (as a more systematics-forgiving cross check on a true autocorrelation), then one is precisely in the regime where the denominator of the B19 estimator has large uncertainties. In Appendix~\ref{sec:nongausserrors} we use some numerical toy models to illustrate how this can cause highly non-Gaussian distributions where our previous propagation-of-error-style error expressions can be misleading. In general, therefore, the conservative approach is to construct a full posterior distribution via sampling.

\section{Simulations} \label{sec:simulations}
Armed with a series of estimators, we now test the limits of their validity. In the following analysis we explore three sets of simulated data (described in the following subsections), all built upon the same simulation and theoretical framework, with additional layers of increasing complexity.  All datasets are composed of a simulated cosmological signal of interest and a noise realization.  Both components have three different scenarios corresponding roughly to a optimistic, conservative, and pessimistic outlook of observational challenges.  In the case of the signals we look at three different theoretical models, beginning with the "perfect" case and then increasing in complexity. For our noise, we assume a Gaussian distribution in all cases, but with increasing variance.

The \emph{absolute} strengths of different high-redshift emission lines are still considerably uncertain, and it is not our goal to make predictions based on a particular physical model.  Since all the estimators discussed here are for a single $k$-mode, we do not require any knowledge of the $k$ dependence of the spectra. For the majority of our analysis, we choose $k=0.45\,h\mathrm{Mpc}^{-1}$ as a representative $k$ mode.

\begin{figure}
\centering
\includegraphics[width=\columnwidth]{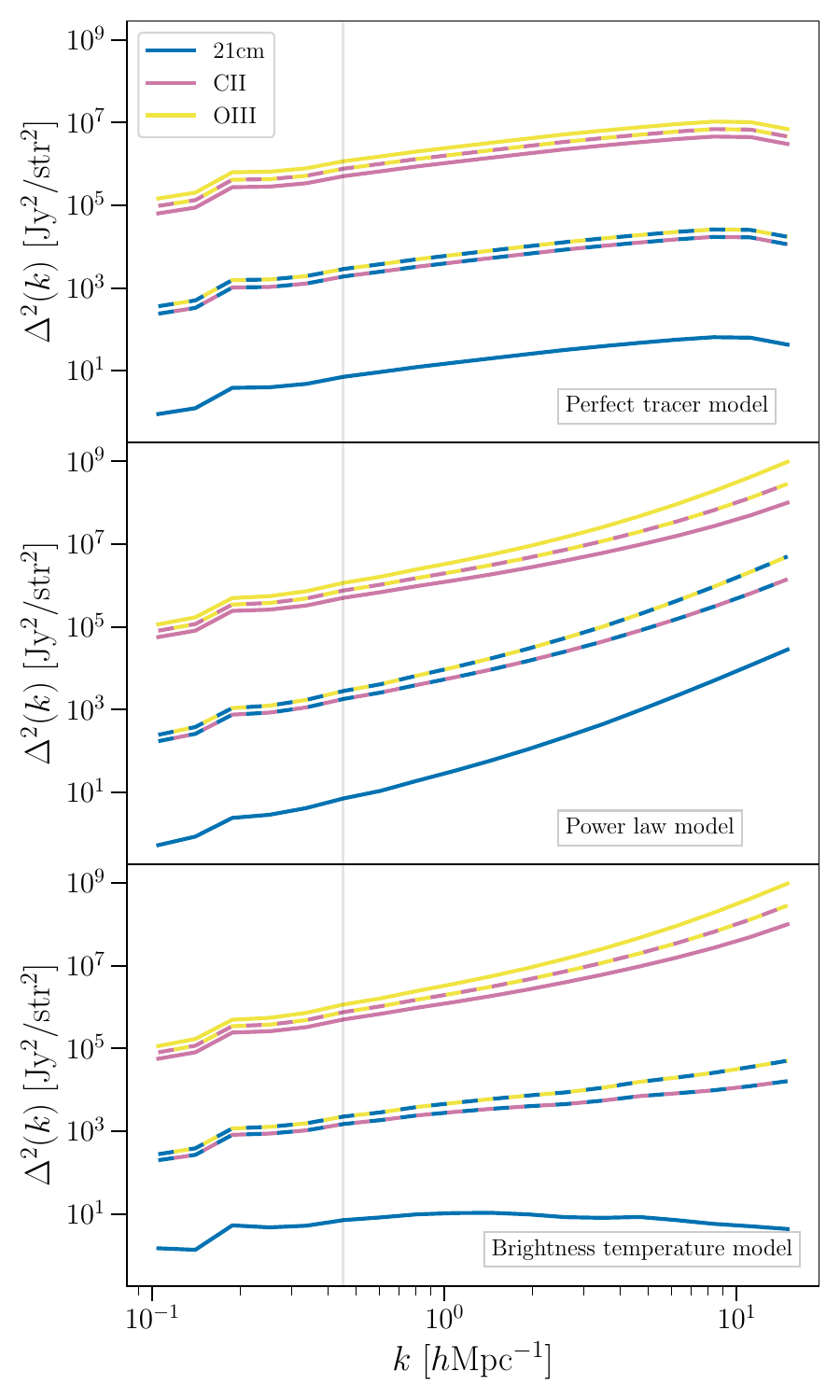}
\caption{Simulated power spectra for the three sets of data considered in our analysis. Top: the perfect tracer model (Section~\ref{sec:sf}) where the data are simulated directly from the system of equations that will eventually be used for our fits. Middle: the power law model (Section~\ref{sec: pl}) where the intensities for all three lines are derived by assuming a power law relationship between the halo masses and the luminosity. Bottom: the bubble model (Section~\ref{sec:bt}) where the 21\,cm brightness temperature is modeled with a more realistic reionization model. Both auto (solid color lines) and cross spectra (dashed color lines) are plotted for each data set, with the alternating colors of the dashed lines indicating the two spectral lines that form the crosscorrelation, e.g., blue (21\,cm) and teal (CII) dashes lines indicate the $P_{21\mathrm{cm/CII}}$ cross spectrum.  The specific $k$ value (vertical light gray) used for the majority of the analysis is also overlaid.}
\label{fig:pspecs}
\end{figure}

However, our analysis is sensitive to the \emph{relative} strengths of each correlation. For all data sets, we fix the amplitude of each autocorrelation to be a reasonable value (described in Section\,\ref{sec: physical}) for each line at the representative $k$ value above.  That amplitude is then fixed to be the same across all three scenarios (separately for each line) by adjusting the normalization of the luminosity relation described in Equation~\eqref{eq:power_law}. This is done to enable a cleaner comparison across the three data sets, and the resulting simulated power spectra for all three data sets are shown in Figure\,\ref{fig:pspecs}, and a summary of the model parameters used are listed in Table\,\ref{tab:vals}.

\begin{figure*}
\centering
\includegraphics[width=\textwidth]{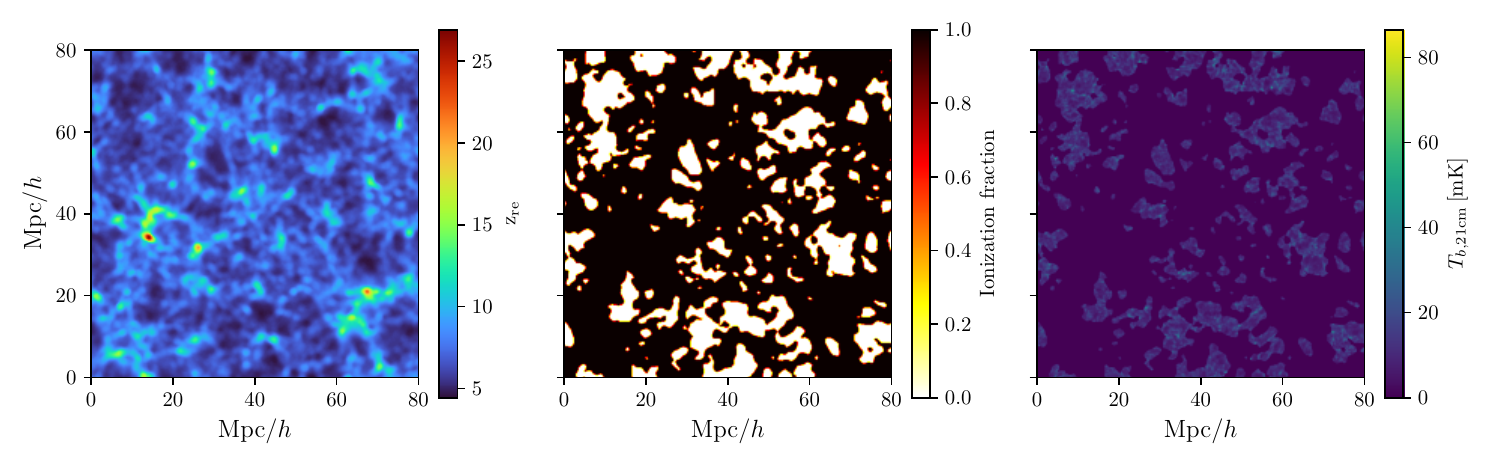}
\caption{2D slices of the simulation products used to generate the brightness temperature field described in Section\,\ref{sec:bt}. The \texttt{zreion} package uses a derived empirical relationship between the density field and the redshift of reionization. In the high spin temperature limit, the redshift at which a given pixel ionizes is related to the density field by a scale-dependent linear bias. This can be used to assign a redshift of ionization for each spatial pixel in the simulation box (left).  From this relation, the ionization field (middle) and 21\,cm brightness temperature field (right) at the desired redshift ($z=6.0155$) can be inferred.}
\label{fig:zreion}
\end{figure*}

 \subsection{Cosmological Signal Models} \label{sec:cosmosigmodel}
We build all three datasets atop an $N$-body dark matter simulation that was run as part of \citet{Mirocha_2021}.  The size of the simulation box is $(80 \, \text{Mpc}/h)^3$, with $512^3$ voxels.  It is run using $2048^3$ DM particles, and is capable of resolving atomic cooling halos (\char`\~$10^8 M_\odot$) with 20 particles. The halos are then identified with a combination of friends-of-friends, and spherical overdensity algorithms.  Once the $N$-body code is run, the halo finder builds a catalog of DM halo masses and their location.  The underlying density field is also saved.  Further detail of the inner workings of the simulation code can be found in \citet{Trac_2015}. The redshift of the simulation box is $z=6.0155$. Lightcone effects \citep{la2014reionization,ghara201521,majumdar2016line,mondal2018towards, datta2014light} are not considered in this work.

With the matter distribution in hand, we describe in the following sections how the intensity maps are constructed in our models.

\subsubsection{Perfect tracer model} \label{sec:sf}
As a jumping off point, our first model follows the linear biasing model outlined in Equation~\eqref{eq:linearbias} exactly. Each spectral line is then a perfect tracer of the matter field.  Each crosscorrelation for this model is created by taking the calculated matter power spectrum from the $N$-body simulation, and simply multiplying it by the appropriate bias factor, as calculated from the autocorrelations, i.e., 
\begin{align}
\beta_i^2 &= \frac{P_{ii}}{P_m}.
\end{align}
This model ignores the shot noise that arises due to the finite number of haloes.

\subsubsection{Power law model} \label{sec: pl}
In this model, the luminosity of each spectral line is related to the underlying halo distribution via a simple power law that relates the luminosity of each halo to its mass,
	\begin{align} \label{eq:power_law}
L_i(M) &= L_{0,i} \left( \frac{M}{M_0} \right)^{\alpha_{i}}
\end{align}
where $L_{0,i}$ and $M_0$ are the characteristic luminosity and mass of the halo respectively, and $\alpha$ sets the slope of the power law. The luminosities are then converted to specific intensities, from which the spectra are then calculated.  The amplitude of each autocorrelation is chosen such that the true bias factor is consistent with the values adopted in B19, except for the 21\,cm amplitude, which is determined by the bubble model below (Section \,\ref{sec:bt}). However, note that this merely fixes the amplitude of the 21\,cm autocorrelation at the reference $k$-value, and the power-law model is used for all three fields in this scenario, including the 21\,cm brightness temperature. 

We are similarly guided in our choice of values for the power index, $\alpha$.  B19 studied three different power law scenarios, where the power law index for each line was set to a low (L), medium (M), or high (H) value. In order to keep our analysis complementary to that of B19, we adopt their L-M power index scenario. In this scenario, we assign the [CII] line a low power index value, $\alpha=0.6$, and give a medium value of $\alpha=1.0$ for [OIII]. The 21\,cm signal is assigned a value of $\alpha=1.3$, which is chosen to be within the range of reasonable values whilst still differing enough from the other lines.
	
\subsubsection{Bubble model} \label{sec:bt}
For our final model, we now model the 21\,cm field in a different fashion than the other spectra lines to account for the fact that at our assumed redshift ($z =6.0155$) there may exist ionized bubbles in the 21\,cm brightness temperature maps. We use the parametric model from \citet{Battaglia_2013} that generates brightness temperature maps based on an inputted  density field.  This parametric model was derived by comparing the relationship between a set of $N$-body simulations of the density field and the redshift of reionization fields. 

To simulate our brightness temperature fields, we use the software library \texttt{zreion}\footnote{Publicly available at \url{https://github.com/plaplant/zreion}.} which takes in a density field and generates a redshift of reionization field based on the empirical prescription of the \citet{Battaglia_2013} model. From the redshift of reionization field, the ionization field can be calculated, and from there the 21\,cm brightness temperature.  We are not particularly interested in the particulars of the parameterization; therefore, we use the fiducial parameter values defined in \citet{Battaglia_2013}.  As a reference, the various fields used to generate the data are shown in Figure\,\ref{fig:zreion}.  

The 21\,cm brightness temperature is then combined with the other spectral lines (formed using the power law mass-luminosity relation outlined above in Section\,\ref{sec: pl}), to form the simulated spectra for this dataset.

\begin{table} 
\caption{Summary of parameter values for the three cosmological signal models described in Section~\ref{sec:cosmosigmodel}. The parameter values are chosen such that each line's autocorrelation power spectrum has the same value across all three simulations when evaluated at the fidicual $k$ used in the analysis. Since the amplitude of the 21~cm signal is derived from the \texttt{zreion} simulation, we report only its intensity.}
\begin{tabular}{lcccc}
\hline
Intensity-weighted & Bias & Intensity & Power Index & Normalization \\
bias $\beta$ & $b$ & $\braket{I}$  & $\alpha$ & $L_0 / M_0^{\alpha}$  \\ 
 & & [\text{kJy}/\text{sr}] & & [$L_{\odot}/M_{\odot}^{\alpha}$] \\
\hline
$\beta_{21\mathrm{cm}} = \braket{T _{21\mathrm{cm}}}$ & --- & 12.4 & 1.3 & $2.6\times 10^{-4}$ \\
$\beta_{\mathrm{CII}} = b_{\mathrm{CII}} \braket{I _{\mathrm{CII}}}$ & 3 & 1.1 & 0.6 & $10.7 \times 10^4$ \\
$\beta_{\mathrm{OIII}} = b_{\mathrm{OIII}} \braket{I_{\mathrm{OIII}}}$ & 5 & 1.00 & 1.0 & $1.15 \times 10^2$  \\
\hline
\end{tabular}
\label{tab:vals}
\end{table}

\subsection{Noise Models}
In addition to the desired cosmological signal, simulated noise is also added to the data.  The type of noise is divided into two categories. The first is a simple fractional percentage of the data signal.  We then model a more sophisticated noise treatment based on both upcoming, and idealized future surveys. The added noise in all cases is Gaussian-distributed.
\subsubsection{Fractional error} A fractional noise percentage on the data (i.e. the crosscorrelations) is assumed.  
The variance $\sigma_{ij}^2$ on a given crosscorrelation $P_{ij}$ are such that $\sigma_{ij} / P_{ij} = [1\%, 10\%, 15\%]$.

\subsubsection{Surveys}\label{sec:surveys} We also consider more realistic, if still idealized, survey configurations.  In this case, the noise is calculated from the experimental parameters of several current or upcoming experiments and takes into the account the reduced sensitivity at higher-$k$ due to the limited resolution. This is often modeled via a window function, $W(k)$, which moderates the true power, $P(k)$, such that the measured power, $P_{\text{meas}}$, is given by
\begin{align} \label{eq: window}
P_{\text{meas}} &= P(k) e^{-k^2 \sigma_{\perp}^2} \int_0^1 e^{-k^2 (\sigma_{\parallel}^2 - \sigma_{\perp}^2) \mu^2} d\mu \equiv P(k) W(k)
\end{align}
The window function is characterized by the hyperparameters, $\sigma_{\perp}$ and $\sigma_{\parallel}$, and contains an integral over the cosine of the spherical polar angle, $\mu \equiv \cos {\theta}$, in Fourier space. The angular resolution of the instrument determines $\sigma_{\perp}$, which is given by
\begin{align}
\sigma_{\perp} &= D(z)\sigma_{\text{beam}}
\end{align}
where $D(z)$ is the comoving distance to the desired redshift, and $\sigma_{\text{beam}}$ is the beam width of the observing instrument.  In the line of sight direction, $\sigma_{\parallel}$ is set by the frequency resolution, such that
\begin{align}
\sigma_{\parallel} &= \frac{c}{H(z)} \frac{\delta_{\nu} (1+z)}{\nu_{\text{obs}}},
\end{align} 
and depends on the spectral resolution $\delta_{\nu}$, the Hubble parameter $H(z)$, and the observing frequency $\nu_{\text{obs}}$ of the instrument. See Appendix C of \citet{Li_2016} for a deeper overview of these expressions.

\begin{figure}
\centering
\includegraphics[width=\columnwidth]{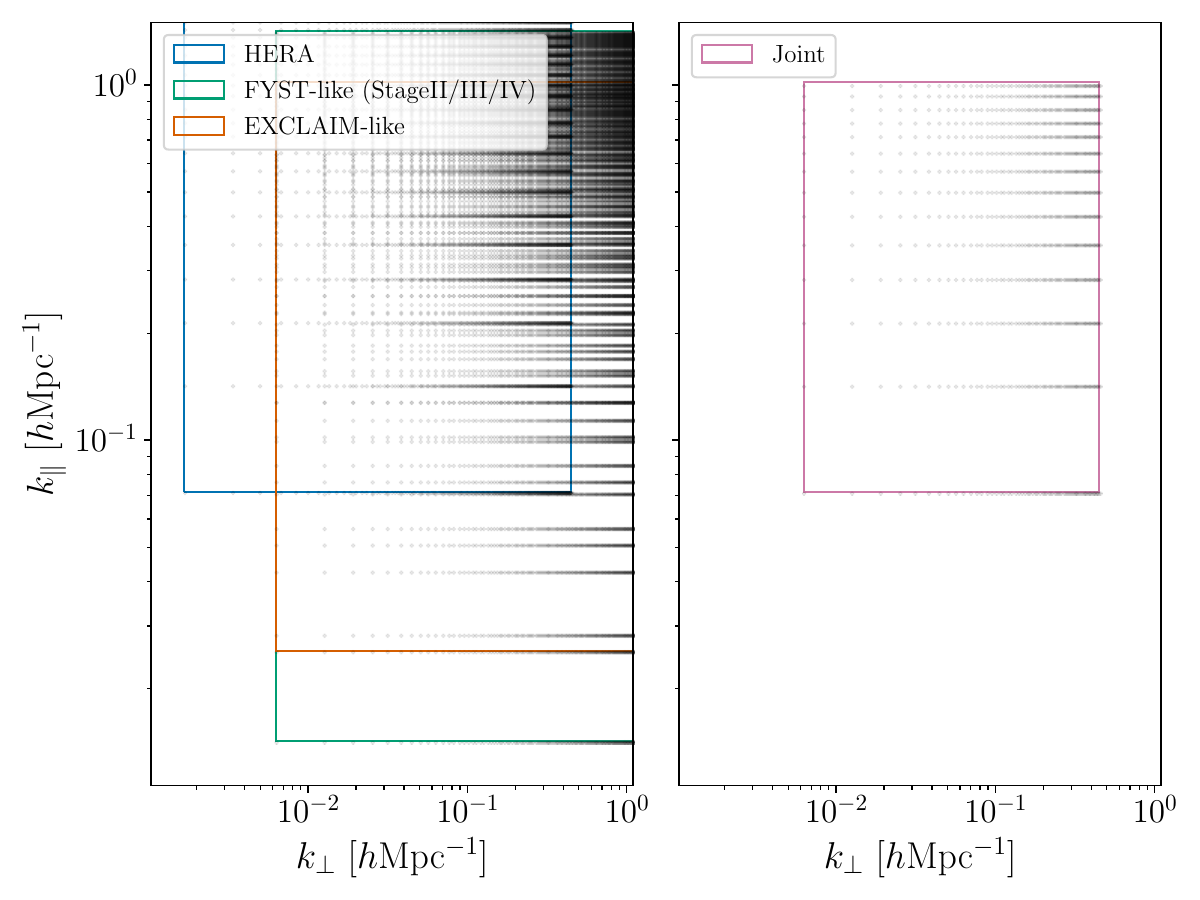}
\caption{Survey coverage in the two-dimensional Fourier space composed of $k_{\parallel}$ and $k_{\perp}$ for an idealized current-generation survey based on the resolution limits of the simulation box.  The native Fourier spacing is shown (black dots) for each independent experiment. It is clear that for the individual survey regions (left panel) the observed $\vec{k}$ are offset from each other. By trimming the surveys to just the overlapping regions in configuration space, the observed $\vec{k}$ will be the same, allowing for the meaningful crosscorrelation of the fields.}
\label{fig:surveys}
\end{figure}

In order to meaningfully cross correlate two fields, they must not only have overlapping survey volumes; they must also overlap in their Fourier space $\mathbf{k}$ coverage. In particular, this must be done at the map level such that each map accesses both the same line-of-sight wavenumbers $k_{\parallel}$ as well as transverse wavenumbers $k_{\perp}$.  In this work, we consider various current or proposed configurations for a suite of LIM experiments. We then modify certain specifications in order to enforce increased complementary coverage in the $k_{\parallel}$-$k_{\perp}$-plane for a set of possible next generation experiments. This greater coverage results in a larger total number of accessible $k$ modes. The $\mathbf{k}$-space coverage by next-generation and futuristic surveys are shown in Figure~\ref{fig:surveys}. While the proposed adjustments to existing experimental configurations are aspirational, they serve to illustrate both the analysis techniques and instrumental requirements in order to measure the parameters to the desired level.

Once the two fields are correlated together, the cross power spectrum is calculated in the usual way. The estimated errors on these spectra are then calculated using the expressions gathered in \citet{Errors}. In particular, the variance $\sigma^2_{ij}$ on a given crosscorrelation measurement is given by\footnote{In principle, there is a minor inconsistency in our treatment of noise. Equation~\eqref{eq:sigmaijwithNmodes} includes contributions from cosmic variance, which should be correlated between different lines (see the discussion in Section~\ref{sec:3cc1p}). However, almost none of our scenarios do we have a cosmic-variance dominated error budget, making this a rather negligible concern.}
\begin{align}\label{eq:sigmaijwithNmodes}
\sigma^2_{ij} &= \frac{(P_{ii} W_{ii} + P_{\text{N},ii})(P_{jj} W_{jj} + P_{\text{N},jj}) + P^2_{ij} W_{ii}W_{jj}}{2 N_{\text{modes}}} 
\end{align}
where $P_{\text{N},ii}$ is the noise power spectrum of the $i$th line (and similarly for $P_{\text{N},jj}$), $W_{ii}$ is the window function [as defined in Equation~\eqref{eq: window}] on the $i$th autocorrelation. The quantity $N_{\mathrm{modes}}$ is the number of independent modes per $k$ bin and is determined by the density of measured $\mathbf{k}$ values in Fourier space. For the [CII] and [OIII] surveys, it is reasonable to think of configuration space as being the native space of observations. The size of the spacing $\Delta k_{i \in \{\parallel, \perp\}}$ of the Fourier grid is then determined by the physical size of the survey region in the line-of sight $(L_{\parallel, \mathrm{surv}})$, and transverse $(L_{\perp, \mathrm{surv}})$ directions, so that
\begin{align}
    \Delta k_{i \in \{\parallel, \perp\}} &= \frac{2\pi}{L_{i \in \{\parallel, \perp\}, \mathrm{surv}}}.
\end{align}
In contrast, most $21\,\textrm{cm}$ surveys are conducted using interferometers, which natively sample Fourier space (at least in the angular directions). The Fourier spacings are then given by
\begin{equation}
    \Delta k_\perp = \frac{2 \pi b_\text{max} }{\lambda D(z)}; \quad
    \Delta k_\parallel = \frac{2 \pi \nu_\text{obs} H(z)}{c (1+z) B_\nu},
\end{equation}
where $B_\nu$ is the total frequency bandwidth, $b_\text{max}$ is the length of the longest baseline, and $\lambda$ is the observing wavelength. See Appendix A of \citet{2014PhRvD..90b3018L} for the derivation of these expressions.

In order to obtain a non-zero cross-correlation, surverys must overlap in both configuration space and Fourier space.  This can be done for real-world surveys by merely trimming down the larger surveys until they fit the physical size of the smallest survey prior to performing the discrete Fourier transform. This then defines the $N_{\mathrm{modes}}$ measured by the set of crosscorrelations. The $k$-modes measured by the current and upcoming set of surveys used in our analysis are shown in Figure~\ref{fig:surveys}.

 Implicit in the inclusion of $N_\text{modes}$ in our expression for $\sigma^2_{ij}$ is an assumption about how Equation~\eqref{eq:P_21} should be evaluated in practice: we are assuming that the individual power spectra are averaged down as much as possible (e.g., in the $\mathbf{k} \rightarrow k$ binning) \emph{before} taking the ratio to compute $\hat{P}_{21}$. This is in contrast to taking the ratio \emph{first} before averaging. In the high signal-to-noise regime, either method will yield the same error bars. This can be seen in Equation~\eqref{eq:errorpropPii}, where the final variance depends linearly on each cross correlation measurement's variance. Therefore, it does not matter whether we average the individual spectra (reducing the right hand side by $N_\text{modes}$) or we average the final result (reducing the left hand side by the same factor). However, this conclusion fails to hold in the low signal-to-noise regime. As shown in Appendix~\ref{sec:nongausserrors}, if the individual cross spectra have insufficiently small error bars, the convergence properties of $\hat{P}_{21}$ become poor and substantial deviations from the truth are possible even with a large number of samples. We therefore recommend an average-then-ratio approach in practice.

The survey volume for an experiment targeting a spectral line with rest wavelength $\lambda_{\text{rest}}$ is given by
\begin{align}
    V_{\text{surv}} = \frac{\lambda_{\text{rest}} (1+z)^2}{H(z)} D^2(z) B_{\nu} S_A.
\end{align}
where $S_A$ is the survey area.

For [CII] and [OIII] observations, the instrumental noise power spectrum is given by
\begin{align}
P_{\mathrm{N},ii} &= V_{\mathrm{pix},ii} \frac{\sigma^2_{\mathrm{N},i}}{t_{\mathrm{pix},ii}}.
\end{align}
where $\sigma^2_{\mathrm{N},i}$ is the noise-equivalent intensity (NEI) of the instrument and $V_{\mathrm{pix},ii}$ and $t_{\mathrm{pix},ii}$ are the volume and observing time per pixel, respectively. 

For the 21~cm signal, we assume an interferometer similar in design to the Hydrogen Epoch of Reionization Array (HERA) as described in \citet{pober2014next}. We use the forecasted signal-to-noise numbers found therein, adapting them for the strength of our simulated $P_{21\mathrm{~cm}}$ signal. For our survey analysis we also assume the same range in $k$-modes that is assumed in their work. Importantly, the treatment in \citet{pober2014next} implicitly assumes that $W(k)$ from Equation~\eqref{eq: window} has already been divided out. We therefore set $W(k) = 1$ for the $21\,\text{cm}$ surveys.

\begin{figure*}
\centering
\includegraphics[width=\textwidth]{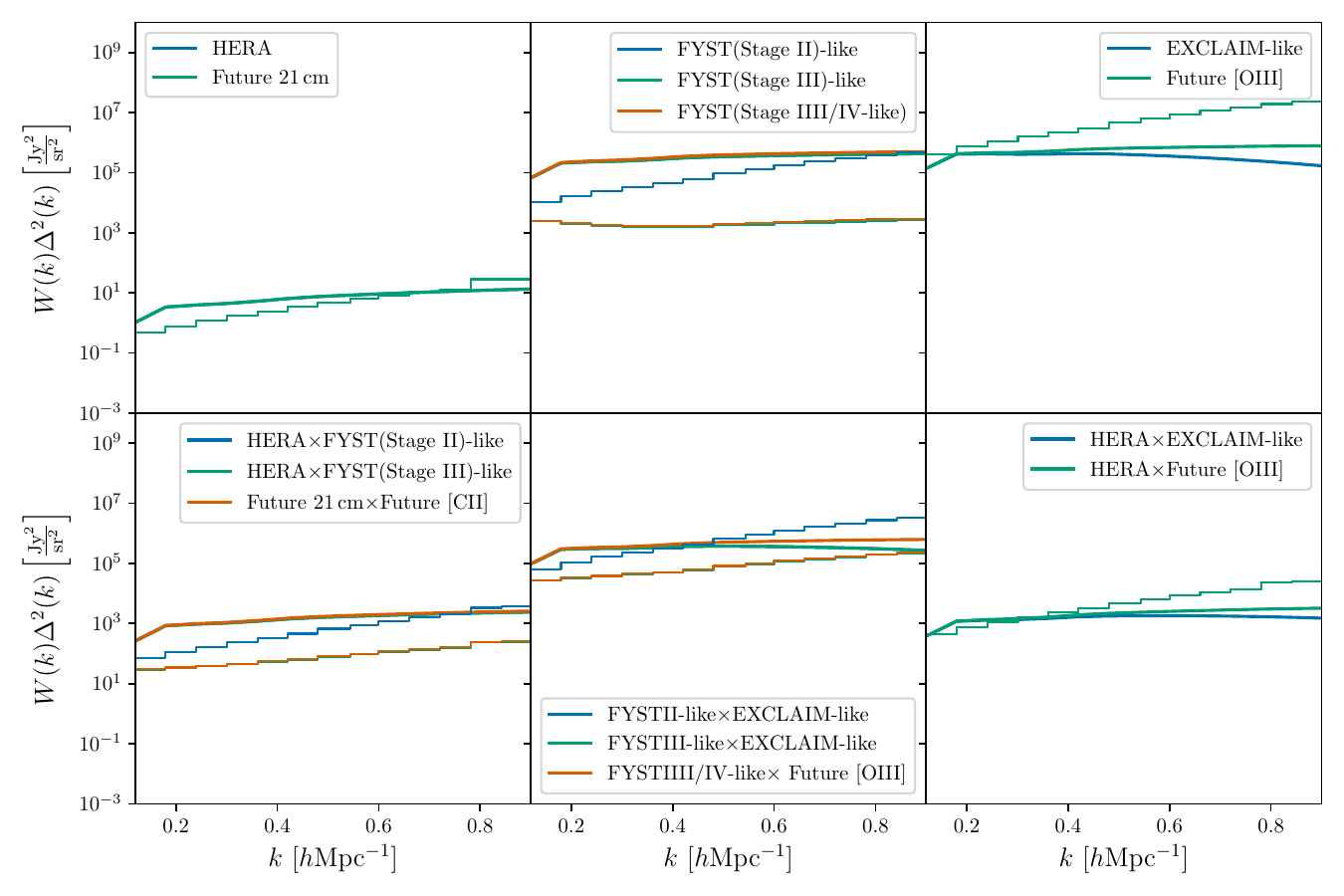}
\caption{Estimated measured signal power spectra in the dimensionless form $ W(k) \Delta^2(k)$ (smooth curves) and associated errors (binned steps) for the current and next-generation surveys discussed in Section~\ref{sec:surveys}. These cosmological signals and error bars are used for the analyses in Section~\ref{sec:survey_forecast}.}
\label{fig:survey_errs}
\end{figure*}

The estimated survey errors of a variety of experimental configuration for all three lines are shown in Figure~\ref{fig:survey_errs}, along with their simulated measured spectra $P_\text{meas}$. These configurations are divided into roughly two categories, current (or upcoming) experiments along with more aspirational future surveys for [CII] and [OIII]. The instrumental parameters for these surveys are listed in Table~\ref{tab:survey_table}, except for the $21\,\text{cm}$ survey, for which we use the errors quoted in \citet{pober2014next} for both the current and future scenarios.

\begin{table*}
	\centering
	\caption{Survey specifications used for the analyses presented in Section~\ref{sec:survey_forecast}. The various specifications are based on the configuration proposals in \citet{Padmanabhan_2020} for the [CII] and [OIII] LIM experiments.  The 21\,cm survey parameters are not listed here since they are based on the HERA experiment errors forecasted in \citet{pober2014next}.}
	\label{tab:survey_table}
	\begin{tabular}{lccccccc} 
		\hline
		Survey & $V_{\text{pix}}$ $[\mathrm{cMpc} / h]^3$ & $\sigma_{\mathrm{N}}^2$ [Jy s$^{1/2} / $sr] & $t_{\mathrm{pix}}$ [s] & $\delta_\nu$ [MHz] & B$_{\nu}$ [GHz] & $S_A [\text{degree}]^2$ & $\sigma_{\mathrm{beam}}$ [arcmin] \\
		\hline
		FYST (Stage II) & 5.96 & $4.84 \times 10^4 $ & 4.06 & 400 & 40.0 & $100.0$ & 0.180 \\
        FYST (Stage III) & 5.96 & $2.10 \times 10^5 $ & $6.5 \times 10^4 $ & 400 & $40.0$ & $100.0$ & 0.180 \\
		EXCLAIM & 381 & $3.00 \times 10^5$ & 201 & 1000 & 40.0 & $100.0$ & 1.22 \\
		Future [CII] & 3.62 & $2.10 \times 10^5$ & $5.26 \times 10^4$ & 1000 & 40.0 & $100.0$ & 0.162 \\
		Future [OIII] & 23.2 & $3.00 \times 10^5$ & 12.2 & 1000 & 40.0 & $100.0$ & 0.300 \\
		\hline
	\end{tabular}
\end{table*}

 \subsection{Simulated Datasets} \label{sec:scenarios}
 In summary, we have three different signal models (i.e. perfect tracer, power law, and bubble), which we then combine with different instrumental noise models (either a fractional error, or one based on survey specifications).  For the bulk of the analysis we work primarily with three scenarios made by combining the signal models and noise levels together to form simulated data sets. We choose three specific combinations intended to investigate qualitatively distinct regimes. These qualitative regimes represent differing degrees of confidence in our modeling and measurements. These scenarios are summarized here.

 \begin{description}
    \item[\textit{The Optimistic Scenario:}] In the simplest case the simulated data is the perfect bias model coupled with low instrumental noise equivalent to a $1\%$ fractional error.  While overly optimistic, it provides a useful reference point for more realistic scenarios.
    \item[\textit{The Conservative Scenario:}] In general, it is not expected that the intensity maps of the universe perfectly follow Equation~\eqref{eq:linearbias}. More realistically, a small amount of tension will exist between lines due to the subtleties in the astrophysics of their origin, resulting in each line tracing the matter distribution in a slightly different way. To better understand both the consequences of this tension, and the technical challenges of intensity mapping measurements, we consider a conservative case of the power law model and higher instrumental noise equivalent to a $10\%$ fractional error.
    \item[\textit{The Pessimistic Scenario:}] The 21\,cm signal is fundamentally different from other spectral lines in its expected distribution with relation to galaxies.  In this scenario, we take the bubble model, with its more sophisticated treatment of the 21\,cm brightness temperature, and combine it with fairly large instrumental noise of $15\%$ fractional error.
\end{description}
The results for these scenarios are presented in Section~\ref{sec:qualitativescenario}. For the concrete survey described in Section~\ref{sec:surveys}, we pair the instrumental noise predictions with the perfect tracer signal model and discuss our results in Section~\ref{sec:survey_forecast}.

\begin{figure}
\centering
\includegraphics[width=\columnwidth]{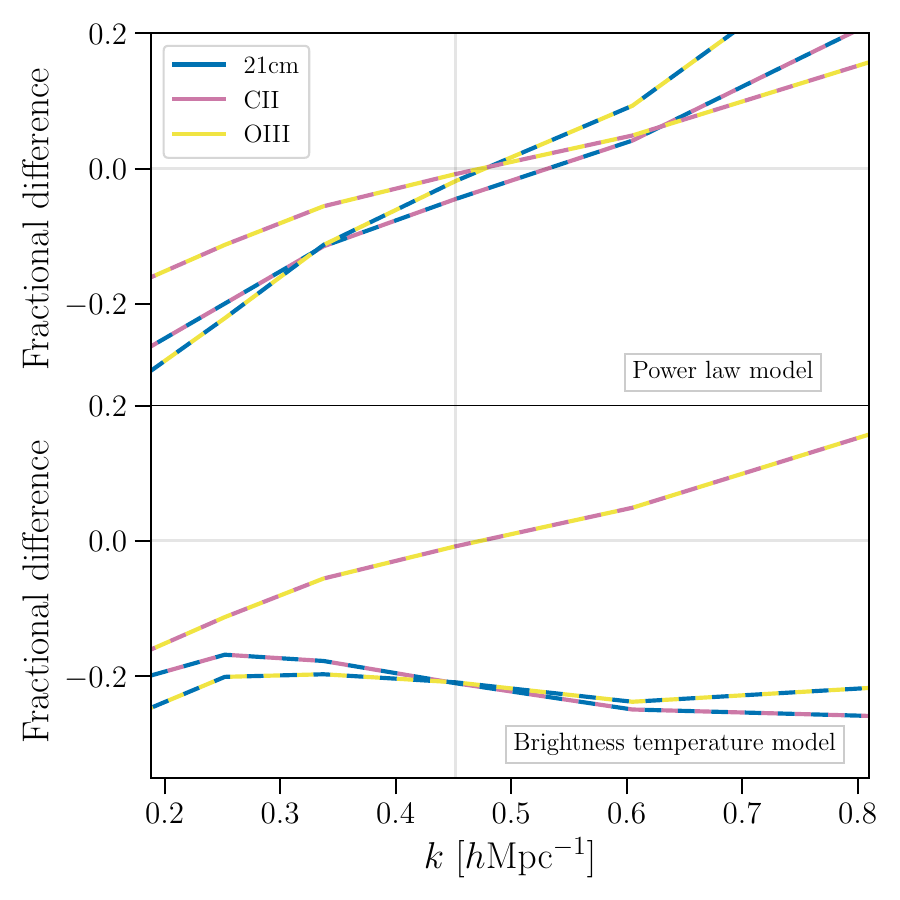}
\caption{
Fractional differences in each cross-correlation power spectrum, comparing the power law and bubble models to the perfect biasing model. The colour scheme follows that of Figure~\ref{fig:pspecs}. The power law model differs less than 5\% at the $k$ value of interest, while the bubble model differs by more than 20\%. Note that the amplitudes of the simulated signals are constructed so that the autocorrelations are identical across signal models at the $k$ value of interest.
}
\label{fig:diffs}
\end{figure}

Note that while each line's autocorrelation has the same amplitude at our fiducial $k$ value by construction when compared across our three scenarios (see Figure~\ref{fig:pspecs}), this does not imply that the same will be true for the cross-correlation spectra. In Figure\,\ref{fig:diffs} we show the fractional difference between the cross-correlation spectra simulated using the power law and bubble models compared to the perfect bias model. The differences are visually evident, and this will have consequences in Section~\ref{sec:qualitativescenario}.

\section{Results} \label{sec:results}
With simulated observations in hand, we now turn to the task of recovering the best-fit values of the parameters.  We first begin by calculating the inferred values from the LSE. We then give our data the full Bayesian treatment as described in Section\,\ref{sec:Bayes!} by using an MCMC algorithm to sample the posterior function. In general, we find that the recovered values of the $P_{21\mathrm{cm}}$ autocorrelation are largely insensitive to the choice of priors on $\beta_{\text{21cm}}$, but highly sensitive to the noise level. We first gain a qualitative understanding by fitting our simulated data based on the three different scenarios described in Section\,\ref{sec:scenarios}, before comparing our results with the projected experimental errors of a suite of current, and next-generation surveys.

\subsection{Qualitative scenarios} \label{sec:qualitativescenario}
A numerical exploration of the posterior distribution was conducted with the three different scenarios outlined in Section\,\ref{sec:scenarios}, using the \texttt{emcee} package \citep{Foreman_Mackey_2013}.  In tandem with the MCMC results we present the best-fit values from our LSE (and therefore B19) estimator.

\begin{figure}
\centering
\includegraphics[width=\columnwidth]{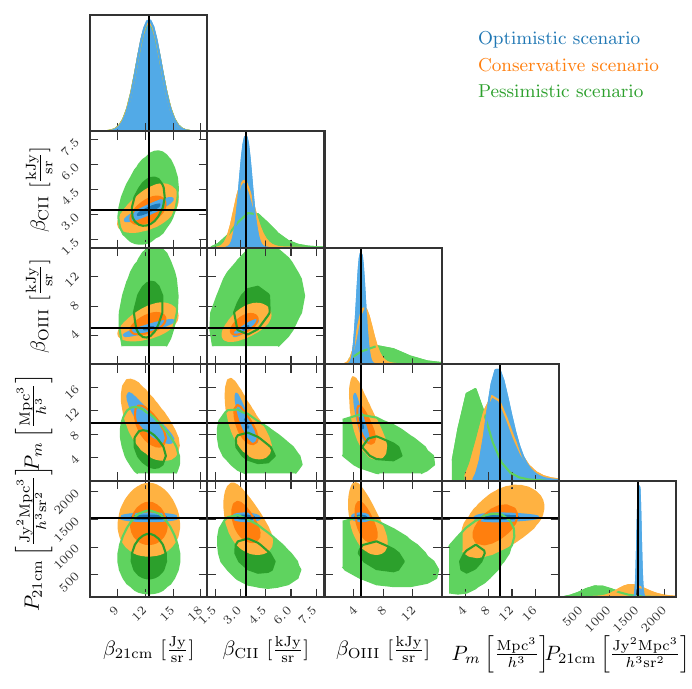} 
\caption{
Posterior distributions for our model parameters and $P_\text{21 cm}$, which is a derived parameter in our analysis. Dark shadings signify $68\%$ credibility regions while light shadings signify $95\%$ credibility. Different colours correspond to the three scenarios defined in Section~\ref{sec:scenarios}. True parameter values are given by the bold black lines. While the optimistic and conservative scenarios produce results consistent with the truth, there is a significant bias with the pessimistic model.}
\label{fig:comp}
\end{figure}

\begin{figure*}
\centering
\includegraphics[width=\textwidth]{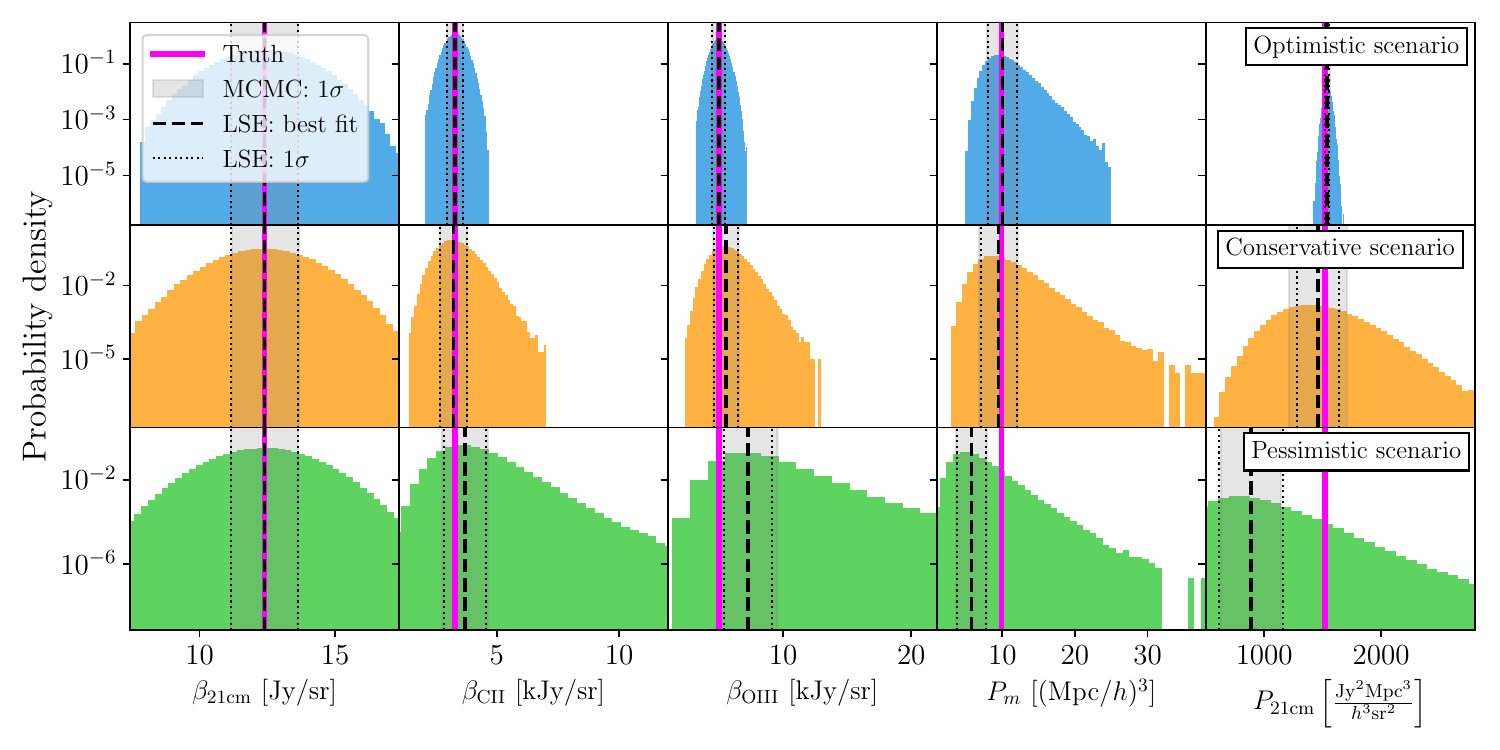} 
\caption{
One-dimensional, marginalized posterior distributions for the four fitted parameters, as well as the reconstructed 21~cm parameter, for all three scenarios described in Section~\ref{sec:cosmosigmodel}. Overlaid are the true values (pink) of the parameters, as well as the $\pm1 \sigma$ range from the MCMC samples (grey shaded), and the best fit values and 1$\sigma$ uncertainties from the LSE (black dashed). Excellent agreement can be seen between the MCMC-predicted uncertainties and those from the LSE formalism. The pessimistic scenario's posterior for $P_\text{21\,cm}$ exhibits a clear skewness. This is due to the cosmological model, not the increased noise compared to the other scenarios.}
\label{fig:estimators_comp}
\end{figure*}

The resultant probability distribution for all three scenarios are shown in Figure\,\ref{fig:comp}. We also zoom in on the marginalized 1D histograms for each scenario in Figure~\ref{fig:estimators_comp}, where we also overlay the results of the LSE formalism. 

In the optimistic regime, the posterior distribution is approximately Gaussian in all parameters save for a long, asymmetric tail for the matter power spectrum parameter, $P_m$. This tail occurs even in the case that the observational errors are initially Gaussian-distributed, and is a consequence of the fact that the matter power density parameter is recovered by taking a ratio of a product, as seen in Equation~\eqref{eq:xhat}. This alters the statistics of the recovered values such that they are not Gaussian-distributed.  In addition to this skewed tail in the matter power spectrum, strong degeneracies exist between the various parameters. This is to be expected given the highly symmetric form of our system of equations. 
 
 While the true value of the matter power spectrum is close to the maximum of the 1D marginalized posterior, it is not equal to the maximum likelihood value (and equivalently the LSE estimator).  The true value is within the 1$\sigma$ range of the sample standard deviation; however, the standard deviation in this case will not fully capture of the uncertainty on the matter power spectrum, due to the skewness of the posterior.

The trend for the conservative scenario is essentially the same, however the widths of the posterior histograms are much wider, indicating the increased uncertainty due to both the imperfect model and elevated noise. In both scenarios the posterior prefers lower values for $P_m$ than the true value, which subsequently causes the value of $P_{21\mathrm{\,cm}}$ to be overestimated.

In our last scenario, we consider a relatively pessimistic scenario which could result in biased parameter recovery. In the power law model, all luminosities for each line are built on top of the dark matter halo catalog.  But in the bubble model, the 21\,cm intensity field is now instead built more directly from the underlying dark matter density field.  While the halo distribution is directly related to the density field, they are not identical in their statistics. Thus in this scenario, we have the [CII] and [OIII] fields still well-correlated with each other and the variance of the underlying halo distribution, while the 21\,cm field is more strongly correlated with the underlying matter density field.

The difference between the statistics of the halo and matter field are sufficient large that this diminishes the correlation between the three lines. This is most clearly seen in Figure~\ref{fig:diffs}, with cross-correlation power spectra involving the 21\,cm line in the bias model having a lower amplitude, signifying some level of decorrelation. The resulting fits are biased away from the true values. From Figure~\ref{fig:comp} we see that in the pessimistic scenario (which uses the bubble model), the matter power spectrum is particularly biased despite being the same for both fields. The final inferred 21\,cm power spectrum also exhibits a bias.

\begin{figure}
\centering
\includegraphics[width=\columnwidth]{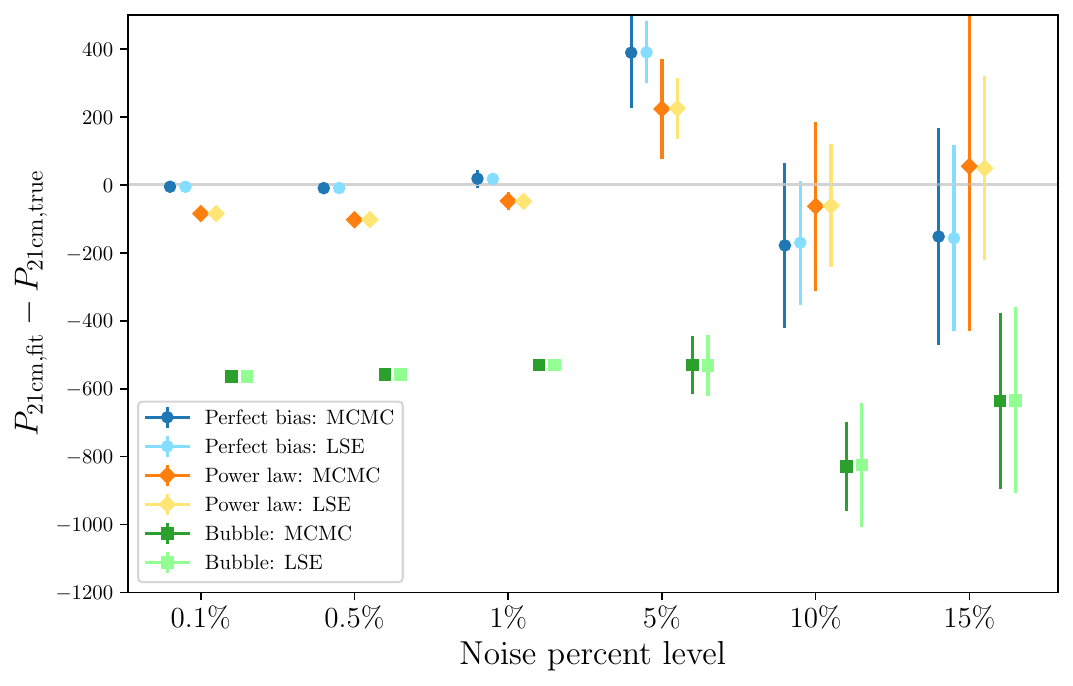}
\caption{Residuals between estimated and true $P_\text{21\,cm}$ as a function of instrumental noise level (expressed as a percentage of the signal). Central values and $1\sigma$ error bars computed from the LSE formalism and the MCMC are found to be largely consistent. In contrast, statistically significant deivations from the truth can occur when there are inconsistencies between the simulated cosmological model and the fundamental assumptions of our method (namely, that all lines trace the matter distribution up to differences in the numerical value of the bias).}
\label{fig:noise_stats}
\end{figure}

In order to better understand the effect of different instrumental noise levels on the recovered $P_{21\mathrm{\,cm}}$, we run a suite of simulations with different fractional crosspower spectrum error bars.  For each noise level, we recover the best fit value and estimated uncertainty of the 21~cm power spectrum from both the MCMC and LSE techniques. The results are shown in Figure~\ref{fig:noise_stats}.  For all noise levels, the LSE results are consistent with the results of the MCMC analysis. Thus, even for relatively large noise levels of $\sim$15\%, the analytic LSE error expression is a good approximation, despite the low noise assumption and first order error propagation.

While the consistency between the two methods is reassuring, the recovered values are not universally good. There is a clear difference between the recovered 21~cm value and the true value when fitting the bubble model, at all noise levels.  This is due to the aforementioned tension between the bubble model and the assumption of the linear biasing model in all our estimators. The fact that the bias exists even at very low noise levels eliminates the possibility that higher noise was the culprit in Figure~\ref{fig:comp}. This low-noise regime also reveals a small inconsistency in the power law case.  While the power law fits are able to recover $P_{21\mathrm{\,cm}}$ within error bars in the higher noise regimes, it starts to show a bias when the noise is low. This is again due to the fundamental tension between the power law and linear biasing models. Granted, this tension is much smaller, as can be seen in the top panel of Figure~\ref{fig:diffs}, and in many noise scenarios the differences are washed out by larger noise. However, for each model there will come a point where further reductions in instrumental noise will not result in an improved measurement of the 21~cm line, simply because of ``systematics" in the theoretical assumptions.

\subsubsection{On the importance of priors}
As noted in Section\,\ref{sec:3cc1p}, it is necessary to assume a prior on at least one bias factor in order to break the degeneracy between the parameters.  For all of the above analysis, a Gaussian prior was assumed on the 21\,cm bias, $\beta_{21\mathrm{\,cm}}$. Conveniently, in the LSE formalism a Gaussian prior on a parameter is mathematically equivalent to an additional, effective data point on said parameter with Gaussian observational errors. 

\begin{figure}
\centering
\includegraphics[width=\columnwidth]{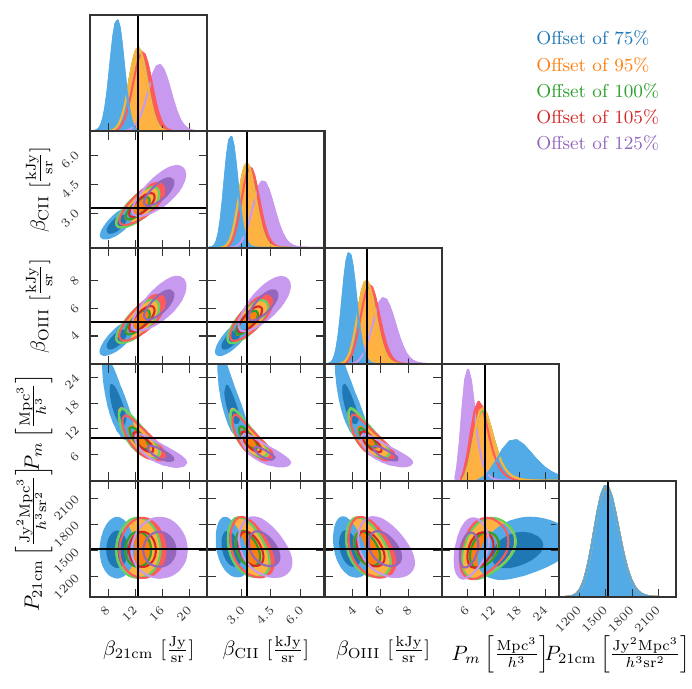}
\caption{Same as Figure~\ref{fig:comp}, but only for the perfect bias signal model and prior distributions on $\beta_\text{21cm}$ that are offset relative to the true value of $\beta_\text{21cm}$ and Gaussian standard deviations that are $10\%$ of the mean values.  While the posteriors for model parameters ($P_m$, $\beta_\text{CII}$, $\beta_\text{21cm}$, and $\beta_\text{OIII}$) are affected by the priors on $\beta_\text{21cm}$, the reconstructed $P_\text{21cm}$ is robust against changes to the prior assumptions (note that the results from all five priors are shown in the bottom right plot; they simply lie exactly on top of each other).}
\label{fig:priors_nl}
\end{figure}

Additionally, during most of our investigation we have centered our Gaussian prior on the true value of the 21\,cm bias factor, which may be unrealistic, since the value of the 21\,cm bias factor is not known to any good precision.  Therefore it is necessary to check that this is not resulting in overly skewed results. To that effect, we conduct an additional noiseless set of MCMCs where the mean of the Gaussian of the prior is offset from the true value. We vary the mean value of the Gaussian prior ranging from $75\%$ to $125\%$ of its true value. In each case, we set the width of the prior to be $10\%$ of the mean value. The results are shown in Figure~\ref{fig:priors_nl}. We see that the choice of prior does not affect the reconstructed 21\,cm power spectrum. This is not surprising given the form of Equation~\eqref{eq:P_21}, where the mean value of the Gaussian prior (i.e. $\beta_0$) is cancelled out in the third step, giving us the B19 estimator which also does not depend on any previous knowledge of the bias parameters. Note, however, that while the choice of prior does not affect the subsequent 21\,cm reconstruction, it does affect the recovered values of the other parameters, even in the noiseless case. This can be a problem particularly for recovering the matter power spectrum. Of course, there is also the additional effect that for all recoveries (including the 21\,cm power spectrum), small biases within expected uncertainties can still occur due to instrumental noise, as seen in Figure\,\ref{fig:comp}. 

\subsection{Idealized survey forecast} \label{sec:survey_forecast}
Finally, we perform numerical experiments corresponding to the concrete survey parameters described in Section~\ref{sec:surveys}, deriving the estimated errors on the crosscorrelation signals from the autocorrelation errors of these surveys. Here, we focus on the $k=0.12$ $h\text{Mpc}^{-1}$ mode since it is one that is forecasted to be able to be measured by HERA to high statistical significance. For plotting convenience we also express our results in this section in terms of dimensionless power spectra $\Delta^2 (k) \equiv k^3 P(k) / 2 \pi^2$.

In our previous analyses, we assumed that the uncertainties on each observation were proportional to the signal; that is to say, the percent error on each line was the same. Therefore, no particular crosscorrelation was noisier than another. One of the critical differences with realistic surveys is that this is no longer the case. Here, some crosscorrelations are measured better than others. For the survey configurations considered herein, the crosspectra between C[II] and O[III], $P_{\mathrm{CII/OIII}}$, is particularly poorly constrained compared to the other two crosscorrelations.

\begin{figure}
\centering
\includegraphics[width=\columnwidth]{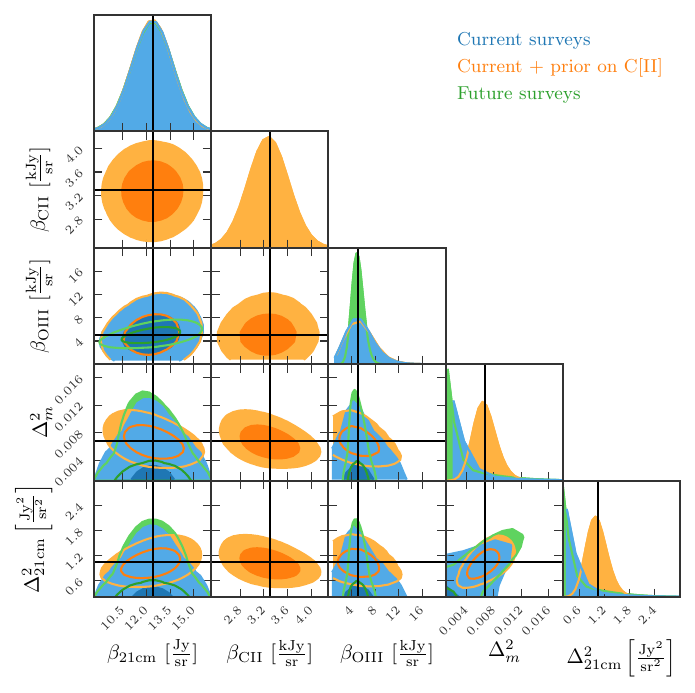}
\caption{Posterior distributions for the three analysis runs conducted on perfect tracer data with no simulated noise, but assuming noise covariance matrices based on the current and future survey configurations outlined in Section~\ref{fig:surveys} (blue and green contours, respectively). All parameters save for $\beta_{\mathrm{21~cm}}$, which has a Gaussian prior, show strong skewness in their histograms.  Additionally, the mean values of the marginalized posteriors are far from the true values for both power spectra and $\beta_{\mathrm{CII}}$. We note that the distributions for $\beta_{\mathrm{CII}}$ in the current and future scenarios are so far from the the true value that they do not appear within the plotted range. However, when an additional prior is added onto a second bias factor, in our case on $\beta_{\mathrm{CII}}$, the resultant posterior (orange) is much better behaved.}
\label{fig:survey_results}
\end{figure}

We fit our perfect tracer simulated data assuming both the current and future noise outlined in Section~\ref{sec:surveys}. The resulting posteriors are shown in Figure~\ref{fig:survey_results} (blue and green contours) for a set of runs with no injected simulated noise.\footnote{In other words, we include the noise effects only in the covariances so as to avoid complicating the interpretation. Without a noise realization injected into our data, any biases from the truth are systematic (for instance due to degeneracies in the fitting process) rather than due to random fluctuations.} For both current and future survey configurations, the uneven noise between the observations results in an extremely broad posterior distribution. In particular, the marginalized posterior for almost all the parameters (sans the one with a prior, a point will we return to in a moment) have long probability tails. In this regime the matter and 21~cm power spectra are poorly constrained, with their posteriors peaking at zero and a large variance (due to the long tails), signifying a non-detection. The preference for $\Delta_m^2$ and $\Delta_\text{21\,cm}^2$ consistent with zero also drives $\beta_\text{CII}$ to very unphysically high values to compensate. (This is in fact why the relevant contours do not appear in the $\beta_\text{CII}$ rows and columns of Figure~\ref{fig:survey_results}---they are off the range of the plots by many, many orders of magnitude). While disappointing, this is unsurprising as it is exactly the scenario considered in the toy model of Appendix~\ref{sec:nongausserrors}.

Fortunately, there is a silver lining here. The fact that $\beta_\text{CII}$ is biased to extremely high values whenever $\Delta_m^2$ and $\Delta_\text{21\,cm}^2$ are poorly constrained suggests a potential mitigation strategy: place a stronger prior on an additional bias parameter, for example $\beta_{\mathrm{CII}}$.  The orange contours in Figure~\ref{fig:survey_results} show the results of an additional analysis where a Gaussian prior is placed on $\beta_{\mathrm{CII}}$. In keeping with our prior on $\beta_{\mathrm{21~cm}}$ we set the standard deviation of the prior to be 10\% of the size of the bias factor. One sees that when this prior is added, the posterior distribution becomes much less pathological and is now able to recover the true value of all parameters. While still exhibiting some small amount of skewness, particularly in the bias factor $\beta_{\mathrm{OIII}}$, the addition of the second prior greatly improves the recovery of the parameters of interest. Therefore, including prior information on at least two bias factors will be critical to recovering the 21~cm signal in this context.

\section{Conclusions} \label{sec: conclusions}
In this work, we build on the estimator proposed in B19 for inferring the 21\,cm power spectrum from measurements of three crosscorrelations between 21cm, [CII], and [OIII]. We ground the B19 estimator in a wider LSE formalism.  This formalism provides an extensible and well-studied framework that can be built upon in a variety of ways, for example by adding additional spectral lines or an autocorrelation measurement. However, the errors forecasted via this method are only valid in the high signal-to-noise limit. Our Bayesian analyses in Section~\ref{sec:results} and our toy models in Appendix~\ref{sec:nongausserrors} reveal that the shapes of the posterior distributions of $\hat{P}_{21}$ can be non-trivial and thus the LSE results must be interpreted carefully. Importantly, uncertainties on the measurement are more complex than can be captured by a simple variance estimate, particularly when the instrumental noise is large.

Varying the complexity of our cosmological models, we find that potential dissonances between the true signal and the linear biasing model can result in biases in parameter inference. This dissonance can be small, as is the case with the power law model (Section~\ref{sec: pl}), or it can be quite large, as is the case for the bubble model (Section~\ref{sec:bt}). In the latter case the trouble arises because some lines directly trace the matter distribution while others trace the matter distribution via dark matter halos. Note that it does not matter whether the different emission lines trace the halo field or the matter field; what matters is for all the lines to trace the same field, regardless of what the field is.

We also conduct a set of more realistic survey analyses, assuming current or potential survey configurations for LIM experiments of 21~cm, [CII], and [OIII]. We find that even for more futuristic proposals, the instrumental noise will be large enough to severely inhibit an unbiased recovery of the 21~cm signal, and indeed the other linear biasing model parameters. However, we also find that the addition of a prior on an additional bias parameter helps to mitigate this issue. Therefore, we emphasize that efforts to better model the biasing of various lines, and thus provide priors for the bias factors, will be critical for the recovery of the 21~cm signal via this method.

In this paper, we have built on intriguing past proposals for constraining auto power spectra from crosscorrelations, creating a systematic mathematical framework for conducting thorough statistical investigations of this idea. Our careful understanding opens the door to rigorous autocorrelation inferences that have the potential to serve as crucial cross-checks on tricky intensity mapping measurements, paving the way towards a comprehensive understanding of our Universe through multiple spectral lines.

\section*{Acknowledgements}
The authors are delighted to thank Sabrina Berger, Patrick Breysse, Dongwoo Chung, Hannah Fronenberg, Ad\'elie Gorce, Jordan Mirocha, Michael Pagano, Bobby Pascua for insightful discussions regarding this work, and Paul La Plante and Jordan Mirocha for sharing their $N$-body simulations. AL and LM acknowledge support from the Trottier Space Institute, the New Frontiers in Research Fund Exploration grant program, the Canadian Institute for Advanced Research (CIFAR) Azrieli Global Scholars program, a Natural Sciences and Engineering Research Council of Canada (NSERC) Discovery Grant and a Discovery Launch Supplement, the Sloan Research Fellowship, and the William Dawson Scholarship at McGill.

\section*{Data Availability}

Code and data used to generate the results in this paper are available upon request.


\bibliographystyle{mnras}
\bibliography{deprism_paper} 



\appendix

\section{Noise covariances for the LSE} \label{sec:LSE_err}
In this Appendix, we derive some of the formal error properties of the LSE discussed in Section~\ref{sec:LSEsubsection}, propagating initial errors in the input cross-correlations to final parameter errors. To make analytic progress, we work in the high signal-to-noise limit while assuming that the errors from different cross-correlations are uncorrelated. At first sight, this may appear to contradict our discussion from Section~\ref{sec:3cc1p}, where we argued that in the cosmic variance-dominated regime, the errors ought to be correlated. However, recognizing that being cosmic-variance dominated is tantamount to having unity signal to noise \emph{per unbinned Fourier mode}, we see that there exists an intermediate regime with high signal to noise in a \emph{binned} power spectrum (assuming statistical isotropy). It is this regime where our analytic expressions hold. Otherwise, one ought to resort to numerical sampling methods such as those described in Section~\ref{sec:Bayes!}.

\subsection{Effect of the linearization of the data set on the statistical properties of the instrumental noise} \label{sec:N}

The first problem to tackle is the logarithm in Equation~\eqref{eq:wdef}. This transforms the noise statistics away from Gaussian. Assuming independence between the different cross-correlations, the noise covariance matrix $\mathbf{N}$ is given by
\begin{align}
\mathbf{N} &= \braket{\mathbf{w w}^T} - \braket{\mathbf{w}} \braket{\mathbf{w}}^T \nonumber \\
&= \begin{pmatrix}
\sigma^2_{w_{ij}} & 0 & 0 & 0 \\
0 & \sigma^2_{w_{jk}} & 0 & 0 \\
0 & 0 & \sigma^2_{w_{ik}} & 0 \\
0 & 0 & 0 & \sigma_{w_{b_i}}^2 \\
\end{pmatrix},
\end{align}
where $\sigma^2_{w_{ij}}$ is the variance on $w_{ij}$ and similarly for the other elements. Consider just the first diagonal element:
\begin{eqnarray}
\sigma^2_{w_{ij}} &=& \braket{w_{ij}^2} - \braket{w_{ij}}^2 \nonumber \\
&=& \left<\left[ \ln \left(1 +  \frac{n_{ij}}{e^{\eta_i} e^{\eta_j} P_m} \right) \right]^2 \right> - \left<\left[ \ln \left(1 +  \frac{n_{ij}}{e^{\eta_i} e^{\eta_j} P_m} \right) \right] \right>^2.
\end{eqnarray}
We can derive an approximation for this expression in the low noise limit (i.e., $ n_{ij} \ll e^{\eta_i} e^{\eta_j} P_m$). Since $\ln{(1+x)} \approx x$ for $x\ll 1$, we have
\begin{eqnarray}
\ln \left(1 +  \frac{n_{ij}}{e^{\eta_i} e^{\eta_j} P_m} \right) \approx \frac{n_{ij}}{e^{\eta_i} e^{\eta_j} P_m} = \frac{n_{ij}}{P_{ij}}.
\end{eqnarray}
The expectation value of this expression is zero if $\langle n_{ij} \rangle = 0$, showing that no noise \emph{bias} is incurred by the logarithm when the signal-to-noise ratio is high (although this is not true in general). Squaring this expression, taking the expectation value, and inserting into $\mathbf{N}$ then gives
\begin{align}
\mathbf{\tilde{N}} = \begin{pmatrix} 
\frac{\sigma_{ij}^2}{P_{ij}^2} & 0 & 0 & 0 \\
0 & \frac{\sigma_{jk}^2}{P_{jk}^2} & 0 & 0 \\
0 & 0 & \frac{\sigma_{ki}^2}{P_{ki}^2} & 0 \\
0 & 0 & 0 & \frac{\sigma_{\beta_i}^2}{\beta_{0}^2},
\end{pmatrix}
\end{align}
which is the approximate noise covariance matrix $\mathbf{\tilde{N}}$ quoted in Equation~\eqref{eq:Ntilde}.

Once we have this form for the noise matrix, we can calculate the projected covariance matrix on the parameters themselves. For notational brevity, we define
\begin{align}
    \xi_{\mu\nu\rho\sigma} \equiv \frac{2\sigma_{\beta_i}^2}{\beta_i^2} + \frac{\sigma_{\mu\nu}^2}{P_{\mu\nu}^2} + \frac{\sigma_{\rho\sigma}^2}{P_{\rho\sigma}^2}  
\end{align}
and  
\begin{align}
    \xi_{i\mu\nu} \equiv\frac{\sigma_{\beta_i}^2}{\beta_i^2} + \frac{\sigma_{\mu\nu}^2}{P_{\mu\nu}^2}
\end{align}
where $\{\mu\nu\rho\sigma\} \in \{i,j,k\}$ and repeated indices are not summed over. Then, the estimated covariance matrix $\boldsymbol{\Sigma}$ on the parameters is
\begin{align} \label{eq:Sigma}
    \boldsymbol{\Sigma} &= [\mathbf{A}^T \mathbf{N}^{-1} \mathbf{A}]^{-1} \nonumber \\
   &= \begin{pmatrix}
  \frac{\sigma_{\beta_i}^2}{\beta_i^2} & \frac{\sigma_{\beta_i}^2}{\beta_i^2} & \frac{\sigma_{\beta_i}^2}{\beta_i^2} & -\frac{2\sigma_{\beta_i}^2}{\beta_i^2} \\
 \frac{\sigma_{\beta_i}^2}{\beta_i^2} & \xi_{ikjk} - \frac{\sigma_{\beta_i}^2}{\beta_i^2}  &  \xi_{jk} &  - \xi_{ikjk} \\
 \frac{\sigma_{\beta_i}^2}{\beta_i^2} &   \xi_{jk} &    \xi_{ijjk}-\frac{\sigma_{\beta_i}^2}{\beta_i^2} &  - \xi_{ijjk} \\
 -\frac{2\sigma_{\beta_i}^2}{\beta_i^2} &  - \xi_{ikjk} &  - \xi_{ijjk} & \frac{\sigma_{\beta_i}^2}{\beta_i^2} + \xi_{ij} + \xi_{jk} + \xi_{ik} \\
   \end{pmatrix}.
\end{align}
\subsection{Recovering the estimated errors on the original parameters} \label{sec:delog}
Next we want to recover the estimator errors on the original parameters, sans logarithm. By exponentiating our errors, we need the uncertainty,$ \sigma_f$, on a function $f(x) = e^{x}$, given the uncertainty on an inferred parameter $x$. For well-behaved functions, this can be found by propagating the error, such that
\begin{align}
    \sigma_f^2 = \left(\frac{\partial f}{\partial x} \right)^2 \sigma_x^2
\end{align}
In our case we have the logarithm of a bias factor, $\eta_i$. The error on the original bias factor $\beta_i$ is what we desire, so $f(\eta_i) = e^{\eta_i} = \beta_i$ and $x=\eta_i$. Then, the propagated error is given by
\begin{align}
\sigma_{\beta_i}^2 &= \beta_i^2 \sigma_{\eta_i}^2,
\end{align}
where $\sigma_{\beta_i}^2$ and $\sigma_{\eta_i}^2$ are the variances on $\beta_i$ and $\eta_i$, respectively. A similar procedure gives us
\begin{align}
\sigma_{P_m}^2 = P_m^2 \sigma_{\ln{P_m}}^2,
\end{align}
with $\sigma_{P_m}^2$ and $\sigma_{\ln{P_m}}^2$ being the variances on $P_m$ and $\ln{P_m}$, respectively.

\subsubsection{Uncertainty on the reconstructed autocorrelation} \label{A:error_prop}
Additionally, we want to estimate the reconstructed errors on an autocorrelation line, $P_{ii} = \beta_i^2 P_m$. Note that we have a function of two parameters, with two different associated uncertainties.  Perturbing the logarithm of $P_{ii}$ gives
\begin{equation}
\delta_{\ln{P_{ii}}} = 2\delta_{\eta_i} + \delta_{\ln P_m },
\end{equation}
where $\delta_{\ln{P_{ii}}}$, $\delta_{\eta_i}$, and $\delta_{\ln P_m }$ are peturbations on $\ln P_{ii}$, $\eta_i$, and $\ln P_m$, respectively. Squaring and taking the ensemble average gives
\begin{eqnarray}
\sigma^2_{\ln{P_{ii}}}  &=& 4 \sigma^2_{\eta_i}  + \sigma^2_{\ln P_m} + 4 \langle \delta_{\eta_i} \delta_{\ln P_m } \rangle \nonumber \\
&=& \frac{\sigma_{ij}^2}{P_{ij}^2} + \frac{\sigma_{jk}^2}{P_{jk}^2} + \frac{\sigma_{ik}^2}{P_{ik}^2},
\end{eqnarray}
where in the last equality we inserted the variances and covariances from Equation~\eqref{eq:Sigma}.
Via the error propagation prescription in Appendix~\ref{sec:delog}, we recover the variance on the autocorrelation, which is given by
\begin{align}
\sigma^2_{P_{ii}}  &= \left( \frac{\sigma_{ij}^2}{P_{ij}^2} + \frac{\sigma_{jk}^2}{P_{jk}^2} + \frac{\sigma_{ik}^2}{P_{ik}^2} \right) P_{ii}^2.
\end{align}

\section{Non-gaussianity and error misestimates} \label{sec:nongausserrors}

In this Appendix, we use simple toy models to numerically demonstrate the regimes where simple error analyses of Equation~\eqref{eq:P_21} break down. Our approach will be to assume that each cross-correlation spectrum in our estimator $\hat{P}_{21}$ is normally distributed about some mean truth. Drawing an ensemble of Monte Carlo realizations then allows the construction of a distribution for $\hat{P}_{21}$.

\subsection{Low-noise denominator}
\begin{figure}
\centering
\includegraphics[width=\columnwidth]{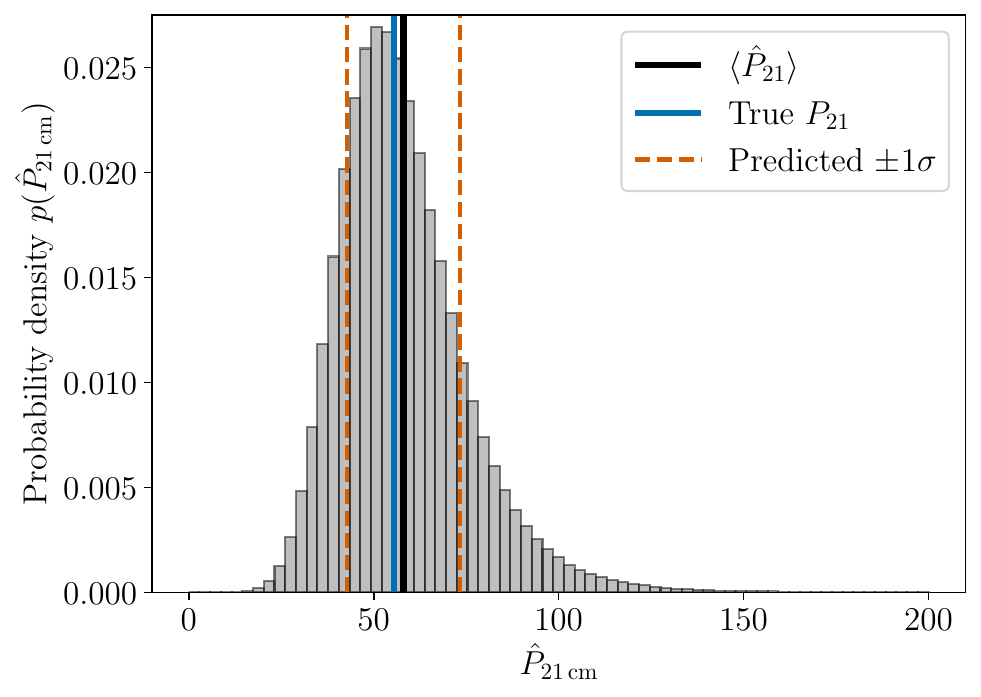}
\caption{A Monte Carlo approximation of the probability density distribution of the recovered $P_{21\mathrm{\,cm}}$ power spectrum from a B19-style estimator in the low noise regime.  Despite some visible skewness, the recovered value if consistent with the true values and within the estimated 1$\sigma$ range.}
\label{fig:good_error}
\end{figure}

Suppose that $P_{21\textrm{cm/CII}}$, $P_{21\textrm{cm/OIII}}$, and $P_{\textrm{CII/OIII}}$ are normally distributed with mean $\mu$ and standard deviation $\sigma$ values of $(\mu, \sigma) = (200.1, 4.0)$, $(5.8, 1.1)$, and $(20.9, 4.18)$, respectively (in some arbitrary units).
We assume that we are not in a cosmic variance-dominated regime so that realizations can be drawn independently from each distribution (see the discussion in Section~\ref{sec:3cc1p}). Doing so and computing $\hat{P}_{21}$ yields Figure~\ref{fig:good_error} for the probability distribution $p(\hat{P}_{21})$. Also shown is the true value of $P_{21}$ as well as the mean value of the Monte Carlo draws and $\pm 1 \sigma$ bounds around this, as predicted by propagation-of-error techniques similar to those used in Appendix~\ref{sec:delog}. Although some a small amount of skewness is evident, the truth is close to the ensemble mean and the simple error estimates are a reasonable summary of the distribution of $p(\hat{P}_{21})$.

\begin{figure}
\centering
\includegraphics[width=\columnwidth]{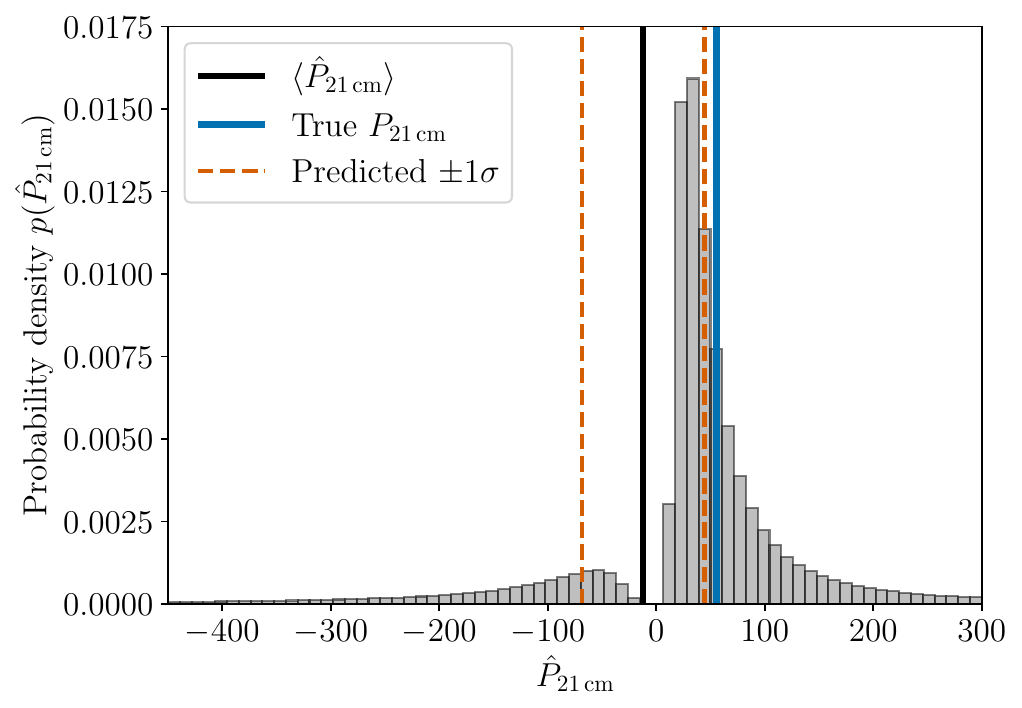}
\caption{Same as Figure~\ref{fig:good_error}, but now assuming a low signal to noise ratio on $P_{\textrm{CII/OIII}}$ itself.  In this case, the probability density is clearly bimodal, and the mean and standard deviation of the samples are biased away from the true value.}
\label{fig:bad_error}
\end{figure}

\begin{figure}
\centering
\includegraphics[width=\columnwidth]{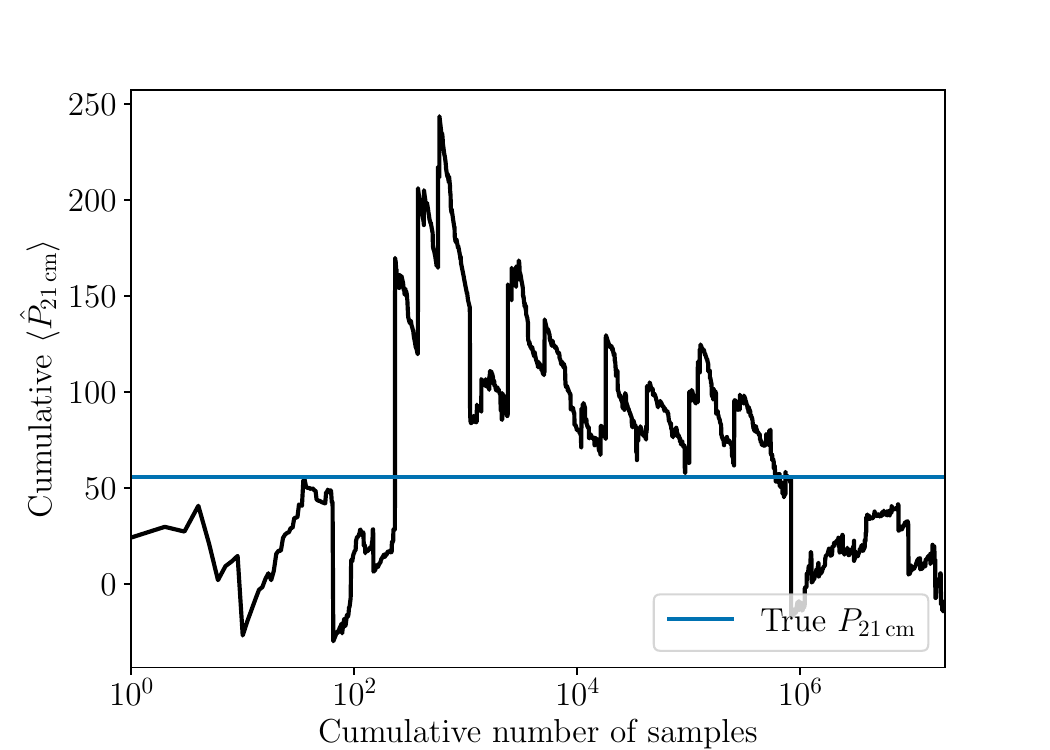}
\caption{The cumulative mean of the recovered $P_{21\mathrm{\,cm}}$ power spectrum from a B19-style estimator as we increase the number of samples. Despite regions of temporary convergence, the resultant mean is easily swayed by outliers.}
\label{fig:cum_P21}
\end{figure}

\subsection{High-noise denominator}
In Figure~\ref{fig:bad_error} we repeat our previous exercise but with a much larger standard deviation for $P_{\textrm{CII/OIII}}$ of $20.9$. This represents a scenario where the $21\,\textrm{cm}$ maps are measured to high significance, but the [CII] and [OIII] maps are not. We immediately see that the probability distribution is no longer unimodal. The low-probability dip near zero is due to the fact that very small values of $\hat{P}_{21}$ are only achievable when $P_{\textrm{CII/OIII}}$ fluctuates to a very large value---an unlikely event. Also evident are the heavy tails of the distribution, which (along with the complicated shape of the distribution) cast doubt on whether the simple error estimates are sufficiently good descriptors of the uncertainties.

The heavy tails of the distribution can also cause the empirical sample mean to be biased, as one can see in Figure~\ref{fig:bad_error}. The exact distance to the truth depends quite sensitively on the random seed, even with a large number of samples. This can be understood by examining the convergence properties of the sample mean. Figure~\ref{fig:cum_P21} shows the cumulative mean as one accumulates randomly drawn samples. It is clear that it is easy to be unlucky and to end up with an estimate that is quite far from the mean. As samples are accumulated, it is easy for temporary converge to be suddenly skewed by outliers. These outliers can have an outsized effect in this context because of the heavy tails of the distribution. As discussed in Section~\ref{sec:surveys}, this motivates averaging down the noise in each individual cross spectrum as far as possible \emph{before} taking the relevant ratio to form $\hat{P}_{21}$, rather than taking the ratio before averaging the result.

\bsp	
\label{lastpage}
\end{document}